\documentclass{article}

\usepackage{arxiv}

\usepackage[utf8]{inputenc} 
\usepackage[T1]{fontenc}    
\usepackage{hyperref}       
\usepackage{url}            
\usepackage{booktabs}       
\usepackage{amsfonts}       
\usepackage{nicefrac}       
\usepackage{microtype}      
\usepackage{lipsum}		
\usepackage{graphicx}
\usepackage{natbib}
\usepackage{doi}
\usepackage{soul}
\usepackage{placeins}
\usepackage{graphicx,psfrag,epsf}
\usepackage{adjustbox}
\usepackage{enumerate}
\usepackage{amsmath}
\usepackage{amsfonts}
\usepackage{amssymb}
\usepackage{amsbsy}
\usepackage[makeroom]{cancel}
\usepackage{bbold}
\usepackage{float}

\usepackage{natbib}
\usepackage{url} 

\usepackage{csvsimple}
\usepackage{comment}

\usepackage{adjustbox}
\usepackage{enumerate}
\usepackage{amsmath}
\usepackage{amsfonts}
\usepackage{amssymb}
\usepackage{amsthm}
\usepackage[makeroom]{cancel}
\usepackage{bbold}
\usepackage{float}
\usepackage[textsize=tiny]{todonotes}

\usepackage[ruled,vlined]{algorithm2e}

\usepackage{url} 
\usepackage{hyperref}
\usepackage{etoolbox}
\usepackage{comment}
\usepackage{array}
\usepackage{bm}

\newtheorem{theorem}{Theorem}
\newtheorem{proposition}{Proposition}
\newtheorem{corollary}{Corollary}
\newtheorem{remark}{Remark}
\newtheorem{lemma}{Lemma}
\newtheorem{condition}{Condition}

\title{A subsampling approach for Bayesian model selection}

\author{ \href{https://orcid.org/0000-0000-0000-0000}{\includegraphics[scale=0.06]{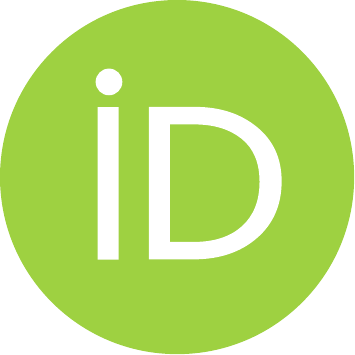}\hspace{1mm}Jon Lachmann}\thanks{The first author, Stockholm University, Stockholm, Sweden. Email: jon.lachmann@stat.su.se} \\
	Department of Statistics\\
	Stockholm University\\
	Stockholm, Sweden\\
	\texttt{jon.lachmann@stat.su.se} \\
		\And
	\href{https://orcid.org/0000-0000-0000-0000}{\includegraphics[scale=0.06]{orcid.pdf}\hspace{1mm}Geir Storvik} \\
	Department of Mathematics\\
	University of Oslo\\
	Oslo, Norway \\
		\And
	\href{https://orcid.org/0000-0000-0000-0000}{\includegraphics[scale=0.06]{orcid.pdf}\hspace{1mm}Florian Frommlet} \\
	CEMSIIS\\
	Medical University of Vienna\\
	Vienna, Austria\\
		\And
	\href{https://orcid.org/0000-0002-3244-6571}{\includegraphics[scale=0.06]{orcid.pdf}\hspace{1mm}Aliaksadr Hubin\thanks{The corresponding author, University of Oslo, 0851 Moltke Moes vei 35 Oslo, Norway. Email: aliaksah@math.uio.no}} \\
	Department of Mathematics\\
	University of Oslo\\
	Oslo, Norway \\
	\texttt{aliaksah@math.uio.no} \\
}

\renewcommand{\shorttitle}{\textit{arXiv} Template}

\hypersetup{
pdftitle={A template for the arxiv style},
pdfsubject={q-bio.NC, q-bio.QM},
pdfauthor={David S.~Hippocampus, Elias D.~Striatum},
pdfkeywords={First keyword, Second keyword, More},
}

\begin{document}
\maketitle

\begin{abstract}
It is common practice to use Laplace approximations to compute  marginal likelihoods in Bayesian versions of generalised linear models (GLM). Marginal likelihoods combined with model priors are then used in different search algorithms to compute the posterior marginal probabilities of models and individual covariates. This allows performing Bayesian model selection and model averaging. For large sample sizes, even the Laplace approximation becomes computationally challenging because the optimisation routine involved needs to evaluate the likelihood on the full set of data in multiple iterations. As a consequence, the algorithm is not scalable for large datasets. To address this problem, we suggest using a version of a popular batch stochastic gradient descent (BSGD) algorithm for estimating the marginal likelihood of a GLM by subsampling from the data. We further combine the algorithm with Markov chain Monte Carlo (MCMC) based methods for Bayesian model selection and provide some theoretical results on convergence of the estimates. Finally, we report results from experiments illustrating  the  performance of the proposed algorithm. 
\end{abstract}

\keywords{Bayesian model selection \and Bayesian model averaging \and Subsampling \and MCMC \and Tall data}

\section{Introduction}

The Marginal Likelihood (MLIK) plays many extremely important roles within Bayesian statistics. For a set of data $\mathbf{y}$ and a model $\mathfrak{m}$ which may depend on some unknown parameters $\bm \theta_\mathfrak{m}$, the marginal likelihood is given by
\begin{equation}
p(\mathbf{y}|\mathfrak{m})=\int_{\bm \Theta_\mathfrak{m}}p(\mathbf{y}|\mathfrak{m},\bm \theta_\mathfrak{m})p(\bm \theta_\mathfrak{m}|\mathfrak{m})d\bm \theta_\mathfrak{m}\label{mlikdef}
\end{equation}
where $p(\bm \theta_\mathfrak{m}|\mathfrak{m})$ is the prior for $\bm \theta_\mathfrak{m}$  under model $\mathfrak{m}$ while $p(\mathbf{y}|\mathfrak{m},\bm \theta_\mathfrak{m})$ is the likelihood function conditional on $\bm \theta_\mathfrak{m}$. 
Consider the comparison of two models $\mathfrak{m}_i$ and $\mathfrak{m}_j$ based on the ratio of their posterior marginal model probabilities:
\begin{equation}
\frac{p(\mathfrak{m}_i|\mathbf{y})}{p(\mathfrak{m}_j|\mathbf{y})} = \frac{p(\mathbf{y}|\mathfrak{m}_i)}{p(\mathbf{y}|\mathfrak{m}_j)}\times \frac{p(\mathfrak{m}_i)}{p(\mathfrak{m}_j)}.\label{BF}
\end{equation}
The ratio between the marginal likelihoods on the right-hand side is called the Bayes Factor~\citep{kass1995bayes}.
It is commonly used to quantify to which extent one model is preferable compared to the other. 
To obtain the ratio~\eqref{BF} the posterior marginal model probabilities need to be calculated up to a constant. 

In the setting of Bayesian model averaging, if we are interested in some quantity $\Delta$ over a given set of models $\mathcal{M}$, we want to calculate the posterior marginal distribution
\begin{equation}\label{posterior_quantile}
p(\Delta|\mathbf{y}) =  \sum_{\mathfrak{m} \in \mathcal{M}}{p(\Delta|\mathfrak{m},\mathbf{y})p(\mathfrak{m}|\mathbf{y})}.
\end{equation}
To this end we need the posterior marginal model probability $p(\mathfrak{m}|\mathbf{y})$ for model $\mathfrak{m}$ which according to Bayes theorem can be calculated as:
\begin{equation}
p(\mathfrak{m}|\mathbf{y}) =  \frac{{p(\mathbf{y}|\mathfrak{m})p(\mathfrak{m})}}{\sum_{\boldsymbol{\mathfrak{m}}' \in\mathcal{M}}{p(\mathbf{y}| \mathfrak{m}')p(\mathfrak{m}')}}.\label{eq:fullpost}
\end{equation}
Again we need to obtain the marginal likelihood $p(\mathbf{y}|\mathfrak{m})$. Further, in the denominator, we would need to compute the marginal likelihood for every model $\mathfrak{m}' \in \mathcal{M}$, something that is usually not computationally feasible except for a very limited model space.
A common solution to this is to use a Metropolis-Hastings algorithm to search through the model space within a Monte Carlo setting~\citep{george1997approaches}. The required acceptance ratios are then of the form
\begin{equation}
r_m({\mathfrak{m}},\mathfrak{m}^{\star})=\min\left\lbrace1,\frac{p(\mathbf{y}|\mathfrak{m}^{\star})p(\mathfrak{m}^{\star})q({\mathfrak{m}}|\mathfrak{m}^{\star})}{p(\mathbf{y}|\mathfrak{m})p(\mathfrak{m})q(\mathfrak{m}^{\star}|{\mathfrak{m}})}\right\rbrace\label{balance}
\end{equation}
and again involve the marginal likelihoods. 

All these problems show the fundamental importance of being able to calculate the marginal likelihood $p(\mathbf{y}|\mathfrak{m})$  in Bayesian statistics. At the same time, it can often be very time consuming to evaluate $p(\mathbf{y}|\mathfrak{m})$, especially when the sample size is large. For this reason, a lot of research has been conducted to find ways to reduce this computational burden. 
When using conjugate priors there exist closed-form expressions for the exact marginal likelihood. Otherwise, it can often be estimated using either simulations or approximation methods~\citep{ando2010bayesian}. These methods range from MCMC simulation to variations of Laplace approximation~\citep{doi:10.1080/01621459.1986.10478240} and {Variational Bayes (VB)}~\citep{friel2011estimating}. Examples of MCMC-based approaches include the harmonic mean estimator~\citep{newtonraftery1994wlb} and Chib's method~\citep{chib1995gibbs}. The former estimates the marginal likelihood as the harmonic mean of a sample obtained using MCMC \citep{friel2011estimating}. The latter is based on the Gibbs sampler~\citep{geman1984gibbs}.

Methods using Laplace approximation of the marginal likelihood are particularly relevant when computational time is of importance \citep{friel2011estimating}, outperforming competing methods in run-time by a large factor. Some rather strict assumptions have to be made about the posterior, but as we shall see later, the performance remains acceptable in most cases, especially when the sample size is large. Yet, even these approximations can become computationally demanding for increasingly large sets of data.
The objective of this paper is to propose a novel subsampling strategy for calculating the marginal likelihood. This will result in more scalable inference for Bayesian model selection. More specifically, the proposed methodology will enable using Bayesian model selection and model averaging for generalized linear models (GLM)  where a large number of  observations is available (so-called \textit{"tall data"}). 

We propose a novel general approach allowing for subsampling optimisation algorithms such as Batch Stochastic Gradient Descent (BSGD) when obtaining Laplace approximation of the marginal likelihood further used in MCMC for Bayesian model selection and averaging. On the implementation part, we use a new variant of the BSGD optimisation algorithm, which we call \textit{Subsampling IRLS with Stochastic Gradient Descent} (S-IRLS-SGD). This algorithm (or standard BSGD) computes the maximum likelihood estimates (MLE) without using the full set of data at every iteration of the optimisation routine. This ultimately allows for a more scalable version of MCMC. The algorithm is available as an R package called \textit{irls.sgd} (\url{https://github.com/jonlachmann/irls.sgd}). We have also created a package called \textit{GMJMCMC} (\url{https://github.com/jonlachmann/GMJMCMC}) which provides an implementation of the Mode Jumping MCMC (MJMCMC) algorithm \citep{hubin2018mjmcmc}. This implementation can be combined with any estimator of the marginal likelihood, including IRLS, SGD, BSGD or S-IRLS-SGD. Both  packages are implemented in the R programming language, but computationally intense parts are written in C++ to additionally enhance scalability. 
These packages provide an implementation of the suggested approach that aims to be ready to use out of the box. Results from extensive simulations as well as real data examples are presented to illustrate the performance of the proposed approach that uses subsampling when estimating MLIKs and MCMC.

The rest of the paper is organised as follows: The class of GLMs is mathematically defined in Section~\ref{section2}. In Section~\ref{sec:subsmcmc}, we introduce the S-IRLS-SGD algorithm used for computing the marginal likelihood and present a general approach on how stochastic gradient optimisation satisfying standard regularity conditions can be combined with MCMC or MJMCMC. In this section, we also provide some theoretical results on the consistency of the algorithm. In Section~\ref{section4}, the proposed algorithm is applied to several simulated and real sets of data. Section~\ref{section5} concludes the paper and gives directions for further research. Additional details and examples are provided in Appendices~\ref{ap:A0}-\ref{A3}. 

\section{Model selection for generalised linear models}\label{section2}

In this section, we will first describe the addressed model space. Then, a description of standard procedures for Bayesian variable selection and their limitations will be presented. To relax some of these limitations, we shall propose a new approach in Section~\ref{sec:subsmcmc}.

\subsection{The model}

We consider the following generalised linear  model: 
\begin{align} \label{themodeleq}
  Y_i|\mu_i,\phi \sim&  \mathfrak{f}(y|\mu_i,\phi),
  \quad \mu_i = \mathsf{h}^{-1}\left(\eta_i\right), \\
   \eta_i =&  \beta_0 + \sum_{j=1}^{p} \gamma_j\beta_{j}x_{ij}.\label{themodeleqend}
\end{align}
Here, $Y_i$ is the response variable while $x_{ij}, j =1,...,p$ are the covariates. We assume $\mathfrak{f}(y|\mu,\phi)$ is a density 
from the exponential family with  corresponding mean $\mu$ and dispersion parameter $\phi$. Applying the link function $\mathsf{h}(\cdot)$ to $\mu_i$, the specific mean of individual $i$, yields $\eta_i$ which is modelled as a linear combination of the covariates. The latent indicators $\gamma_j\in\{0,1\}, j = 1,...,p$ define if covariate $x_{ij}$ is  to be included in the model ($\gamma_j = 1$) or not ($\gamma_j = 0$) while
$\beta_j \in \mathbb{R}, j=0,...,p$ are the corresponding regression coefficients. Note that, in the Introduction as well as in other sections we use the common notation that when we condition on $\mathbf{y}=(y_1,...,y_n)$ we implicitly also condition on $\{x_{ij}\}, i = {1,...,n}, j = 1,...,p$.

To put the model into a Bayesian framework we have to assign prior distributions to all parameters. Concerning the model structure, we specify
\begin{align}
\gamma_j|q \sim& \text{Binom}(1,q),\quad j=1,...,p\label{glmgammaprior}
\end{align}
where $q$ is the prior probability of including a covariate into the model. For the sake of simplicity, prior distributions for different $\gamma_j$'s are assumed to be independent and identical.  The specific choice of priors for $\boldsymbol{\beta}$ and $\phi$ for different applications will be discussed in Section~\ref{section4}. 

\subsection{MCMC for Bayesian variable selection}
Given a set of $ p $ covariates, we can define a specific model through the vector $ \mathfrak{m} = (\gamma_1,...,\gamma_p)$. Following the notation used in the Introduction, denote the model space containing $ 2^p $ possible models as $ \mathcal{M} $. Also for the time being, assume the marginal likelihoods $p(\mathbf{y} | \mathfrak{m})$ are available.

If a full enumeration of every model is possible, one can compute the exact model posteriors through Equation~\eqref{eq:fullpost}. Alternatively, estimates of the posterior probabilities can be obtained through the use of an MCMC algorithm. We can for example construct a Metropolis-Hastings algorithm by applying the acceptance probabilities from Equation~\eqref{balance}. There, we introduce some proposal distribution $ q(\mathfrak{m}^\star | \mathfrak{m}) $ that works on the marginal space of models.
The standard \textit{MCMC estimator} (MC) of the posterior after $ t $ samples $ \mathfrak{m}^{(s)}, s = 1,...,t $ is then
\begin{align}
	\hat p_{MC}^{(t)} (\mathfrak{m} | \mathbf{y}) = t^{-1}\sum_{s=1}^{t} \mathbb{1} (\mathfrak{m}^{(s)} = \mathfrak{m}).
\end{align}
Alternatively, the \textit{renormalized estimator (RM)}~\citet{clyde1996} of the marginal posterior is given by
\begin{align}
	\hat p_{RM}^{(t)} (\mathfrak{m} | \mathbf{y}) = \frac{p(\mathbf{y} | \mathfrak{m}) p(\mathfrak{m})}{\sum_{\mathfrak{m}' \in \mathcal{M}_V^{(t)}} p(\mathbf{y} | \mathfrak{m}') p(\mathfrak{m}')} \mathbb{1}(\mathfrak{m} \in \mathcal{M}_V^{(t)}), \label{renorm}
\end{align}
where $ \mathcal{M}_V^{(t)} $ denotes the set of all unique models visited during a MCMC run of $t$ iterations. This alternative estimator utilises the availability of the marginal likelihood $p(\mathbf{y} | \mathfrak{m})$.
The estimates from both of the estimators will, under standard assumptions, converge to the true posterior as the number of iterations $T \to \infty $, but the RM estimates will typically converge faster~\citep{clyde1996,hubin2018mjmcmc}. 

It is straightforward to see that based on these two estimators we can obtain estimators for the inclusion probability of any given variable as
\begin{align}
	\hat p_{MC}^{(t)}(\gamma_j = 1 | \mathbf{y}) &=\sum_{\mathfrak{m} \in \mathcal{M}_V^{(t)}} \mathbb{1} (\gamma_j = 1) \hat{p}_{MC}^{(t)}(\mathfrak{m} | \mathbf{y}) \label{inclusion1}\\
	\intertext{and}
	\hat p_{RM}^{(t)}(\gamma_j = 1 | \mathbf{y}) &= \sum_{\mathfrak{m} \in \mathcal{M}_V^{(t)}} \mathbb{1} (\gamma_j = 1) \hat{p}_{RM}^{(t)}(\mathfrak{m} | \mathbf{y}). \label{inclusion2}
\end{align}

\subsubsection{Mode Jumping MCMC (MJMCMC)} \label{section:mjmcmc}

The distribution of models will often be multi-modal.
Sampling from a complicated multi-modal distribution using standard MCMC algorithms, such as the Metropolis-Hastings algorithms described in the previous subsection, often leads to problems. A proposal distribution needs to be chosen in a way that allows for both thorough exploration of a nearby mode, as well as the possibility to make transitions to other modes in the parameter space being explored. Choosing a proposal that makes mostly small steps will explore the current location thoroughly, but almost never escape it. On the other hand, a proposal that suggests large jumps will very seldom find the large jumps to be accepted since the algorithm will end up in points of low probability with respect to the target distribution being explored.

To solve this problem, \citet{tjelmelandhegstad2001mjmcmc} introduced \textit{Mode Jumping proposals}, creating the MJMCMC algorithm for continuous variables.
\citet{hubin2018mjmcmc} later adapted it to work for discrete binary variables, used in the case of model selection.
Assume that the current state is $\mathfrak{m}$, proposals are then generated in three steps. First, a large jump is made by swapping a large proportion of the binary variables $ \gamma_j\rightarrow 1-\gamma_j$, giving $\mathfrak{m}_0^{\star}$. Second a local optimisation is performed \citep[based on theory on combinatorial optimisation,][]{blumroli2003meta}, giving $\mathfrak{m}_1^{\star}$. Finally, a small perturbation of the local optimum is made by swapping a small proportion of the binary variables, giving $\mathfrak{m}^{\star}$. 
To be able to calculate the acceptance probability, two backwards intermediate states are also calculated to get the reverse path $ \mathfrak{m}^{\star} \to \mathfrak{m}_0 \to \mathfrak{m}_1 \to \mathfrak{m} $.  Following the description from \citet{hubin2018mjmcmc}, this process is described in more detail in Algorithm~\ref{jump_mjmcmc}.
A proof of convergence was given in~\citet{tjelmelandhegstad2001mjmcmc} for the continuous case and adapted to the discrete case in~\citet{hubin2018mjmcmc}. 

\begin{algorithm}[!ht]
	\SetAlgoLined
	Sample a large set of components $I\subset\{1,...,p\}$ uniformly without replacement.\\
	Generate $\mathfrak{m}^{\star}_{0}$ by swapping the components $I$ of $\mathfrak{m}$.\\
		Perform the local optimisation defined as $ \mathfrak{m}^{\star}_1 \sim q_{o}(\mathfrak{m}^{\star}_1 | \mathfrak{m}^{\star}_{0}) $.\\
		Perform a small randomisation to generate the proposal $ \mathfrak{m}^{\star} \sim q_r(\mathfrak{m}^{\star} | \mathfrak{m}^{\star}_1) $.\\
		Reverse the large jump by swapping the components $I$ of $\mathfrak{m}^{\star}$.\\
		Do the  backwards optimisation $ \mathfrak{m}_1 \sim q_l(\mathfrak{m}_1 | \mathfrak{m}_{0}) $.\\
		Calculate the probability of doing a small randomisation to the starting value $ q_r(\mathfrak{m}|\mathfrak{m}_1) $.\\
		Set the next state of the chain to\\
		\begin{align}
			\mathfrak{m}' = \begin{cases}
				\mathfrak{m}^{\star} & \text{with probability } \alpha = r^{\star}_{mh}(\mathfrak{m}, \mathfrak{m}^{\star}; \mathfrak{m}_1, \mathfrak{m}_1^{\star})\\
				\mathfrak{m} & \text{otherwise,}
			\end{cases}
		\end{align}
	 	where
		\begin{align}
			r^{\star}_{mh}(\mathfrak{m}, \mathfrak{m}^{\star}; \mathfrak{m}_1, \mathfrak{m}_1^{\star}) = \min \left\lbrace 1, \frac{\pi(\mathfrak{m}^{\star}) q_r (\mathfrak{m} | \mathfrak{m}_1)}{\pi(\mathfrak{m}) q_r(\mathfrak{m}^{\star} | \mathfrak{m}^{\star}_1)} \right\rbrace.
		\end{align}
	\caption{Mode Jumping proposal algorithm}
	\label{jump_mjmcmc}
\end{algorithm}

\citet{tjelmelandhegstad2001mjmcmc} note that the mode jumping proposals should only be used a fraction of the time, the rest of the proposals are generated through regular Metropolis-Hastings kernels. They also demonstrate that the algorithm is efficient at exploring complicated target distributions with multiple modes by various examples. 

\subsection{Laplace approximation of the marginal likelihood} \label{section:laplace}

So far we have assumed that the marginal likelihoods $p(\mathbf{y} | \mathfrak{m})$ are available. This is however rarely the case in practice. The model prior $p(\mathfrak{m})$ is often trivial to calculate, but the marginal likelihood \eqref{mlikdef} is generally less tractable. Computationally expensive calculations of marginal likelihoods become particularly problematic for model selection, where a large number of marginal likelihoods needs to be calculated. 
For this reason, approximations are widely used, most of which are building on asymptotic properties of the posterior \citep{ando2010bayesian}. A prime example is the Laplace approximation, which is central to our subsampling approach.

The Laplace method  \citep{laplace1986} makes use of the Taylor  approximation to obtain approximate solutions of integrals.
It is common to apply the Laplace approximation for the likelihood multiplied by the prior for the model parameters $p(\mathbf{y} | \bm \theta_\mathfrak{m},\mathfrak{m})  p(\bm \theta_\mathfrak{m})$. Let $ \bm{\tilde{\theta}}_\mathfrak{m} $ denote the posterior mode. Then, the approximation of the marginal likelihood looks as follows:
\begin{align}
	\hat{p}(\mathbf{y} | \mathfrak{m}) = p(\mathbf{y} | \bm{\tilde{\theta}_\mathfrak{m}},\mathfrak{m}) p(\bm{\tilde{\theta}}_\mathfrak{m}) (2 \pi)^{|\bm \theta_\mathfrak{m}|/2}\left|S(\bm{\tilde{\theta}_\mathfrak{m}},\mathfrak{m}) \right|^{-1/2}, \label{laplace}
\end{align}
where $ S (\bm{\tilde{\theta}}_\mathfrak{m},\mathfrak{m})$ is the Hessian of $ \ln [p(\mathbf{y} | \bm{\tilde{\theta}}_\mathfrak{m},\mathfrak{m})  p(\bm {\tilde \theta}_\mathfrak{m})] $ evaluated at the posterior mode $\bm{\tilde{\theta}}_\mathfrak{m}$ \citep{ando2010bayesian}.

For most natural priors one can disregard its effect on the posterior for large $ n $. This yields the \textit{Bayesian Information Criterion (BIC)} \citep{schwarz1978bic} and a computationally efficient approximation of the marginal posterior. One uses the MLE $ \bm {\hat\theta}_\mathfrak{m} $ instead of the posterior mode in \eqref{laplace} and the Hessian $ S(\bm{\tilde{\theta}, \mathfrak{m}}_\mathfrak{m}) $ is also replaced by the observed Fisher information matrix
\begin{align}
	J_n(\bm{\hat{\theta}}_\mathfrak{m},\mathfrak{m}) = - \left. \nabla^2  \ln p(\mathbf{y} | \bm \theta_\mathfrak{m},\mathfrak{m})\right|_{\bm \theta_\mathfrak{m} = \bm{\hat \theta}_\mathfrak{m}}\; .
\end{align}
Taking the logarithm and removing terms that for large $ n $ will be negligible, we get the approximation of the log marginal likelihood
\begin{align} 
	\ln p(\mathbf{y} | \mathfrak{m}) \approx \ln p(\mathbf{y} | \bm{\hat{\theta}}_\mathfrak{m},\mathfrak{m}) - \frac{|\bm \theta_\mathfrak{m}|}{2} \ln n. \label{margpost}
\end{align}
Asymptotically, \eqref{laplace} and \eqref{margpost} are equivalent, in the following, we will use the latter.
The approximation~\eqref{laplace} becomes exact for a model with Gaussian observations and a normal prior for the coefficients $\bm \beta$ since the log of the likelihood for the observations is then taking a negative quadratic form; otherwise, it would have an error of order $\mathcal{O}(n^{-1})$. 

\section{Subsampling for Bayesian model selection}\label{sec:subsmcmc}

Combining Equation~\eqref{renorm} with Laplace approximations  $\hat p(\mathbf{y}|\mathfrak{m})$  from Equation \eqref{laplace}, we get  
\begin{align}
\hat p(\mathfrak{m}|\mathbf{y})=\frac{p(\mathfrak{m})\hat p(\mathbf{y}|\mathfrak{m})}{\sum_{\mathfrak{m'}\in\mathcal{M}}p(\mathfrak{m'})\hat p(\mathbf{y}|\mathfrak{m'})}.\label{renormlaplace}
\end{align}
In this paper we are primarily concerned with the situation where $n$ is very large. In that case the approximation~\eqref{margpost} is expected to work well for most priors and also for the likelihood of the most common models.  For Gaussian likelihoods with conjugate priors, the result is exact, $\hat p(\mathfrak{m}|\mathbf{y}) = p(\mathfrak{m}|\mathbf{y})$ as $\hat p(\mathbf{y}|\mathfrak{m})= p(\mathbf{y}|\mathfrak{m})$. The approximation will be suitable for more general GLMs, since as $n$ increases, we have $\hat p(\mathbf{y}|\mathfrak{m})\rightarrow p(\mathbf{y}|\mathfrak{m})$. We quickly show this in the following straight-forward Proposition~\ref{prop1}.
\begin{proposition}\label{prop1}
If $\hat p(\mathbf{y}|\mathfrak{m})\rightarrow p(\mathbf{y}|\mathfrak{m})$ as $n\rightarrow\infty$ and $\sum_{\mathfrak{m'}\in\mathcal{M}}p(\mathfrak{m'}) = 1$ then $\hat p(\mathfrak{m}|\mathbf{y})\rightarrow p(\mathfrak{m}|\mathbf{y})$ as $n\rightarrow\infty$.
\begin{proof}
By taking the limit of the left and right sides of Equation~\eqref{renormlaplace}, we get:
\begin{align*}
    &\lim_{n\rightarrow\infty}\hat p(\mathfrak{m}|\mathbf{y}) = \lim_{n\rightarrow\infty}\frac{p(\mathfrak{m})\hat p(\mathbf{y}|\mathfrak{m})}{\sum_{\mathfrak{m'}\in\mathcal{M}}p(\mathfrak{m'})\hat p(\mathbf{y}|\mathfrak{m'})}\\
    &= \frac{p(\mathfrak{m})\lim_{n\rightarrow\infty}\hat p(\mathbf{y}|\mathfrak{m})}{\sum_{\mathfrak{m'}\in\mathcal{M}}p(\mathfrak{m'})\lim_{n\rightarrow\infty}\hat p(\mathbf{y}|\mathfrak{m'})} =  \frac{{p(\mathbf{y}|\mathfrak{m})p(\mathfrak{m})}}{\sum_{\boldsymbol{\mathfrak{m}}' \in\mathcal{M}}{p(\mathbf{y}| \mathfrak{m}')p(\mathfrak{m}')}} = p(\mathfrak{m}|\mathbf{y}).
\end{align*}
\end{proof}
\end{proposition}

However, as $n$ increases, the computation complexity required to obtain the MLE does so as well. This is the primary motivation for the subsampling approach we introduce next.

The suggested approach can be framed into a general idea of simulating from a time-inhomogeneous MCMC (TIMCMC) algorithm  \citep{douc2004quantitative} allowing for the theory developed for TIMCMC to hold \citep{douc2004quantitative, saloff2011merging, fort2011convergence}. TIMCMCs for each time-step have the following target 
\begin{align}
\hat p^{(t)}(\mathfrak{m}|\mathbf{y})=\frac{p(\mathfrak{m})\hat p^{(t)}(\mathbf{y}|\mathfrak{m})}{\sum_{\mathfrak{m'}}p(\mathfrak{m'})\hat p^{(t)}(\mathbf{y}|\mathfrak{m'})}.
\end{align}
Here, $\hat p^{(t)}(\mathbf{y}|\mathfrak{m})$ is some estimate of $\hat p(\mathbf{y}|\mathfrak{m})$ at time $t$. In the rest of the section, we shall construct $\hat p^{(t)}(\mathbf{y}|\mathfrak{m})$ through an adaptation of a stochastic gradient optimisation such that
$\hat p^{(t)}(\mathbf{y}|\mathfrak{m})$ converges to $\hat p(\mathbf{y}|\mathfrak{m})$ as $t$ increases.




\subsection{
Estimating the  MLIK for GLMs with stochastic gradient optimisation}


Maximum likelihood estimation (MLE) within the context of GLM is usually performed via \textit{Iterative Reweighted Least Squares (IRLS)}  \citep{beaton1974irls, schlossmacher1973irls}, where a \textit{Weighted Least Squares (WLS)} equation  is iteratively reweighted with the inverse of  residuals from the previous solutions. 
In general, the most well-known optimisation algorithm exploiting a subsampling approach is \textit{Stochastic Gradient Descent (SGD)}. According to \citet{robbinsmonro1951sgd}, the SGD optimisation algorithm asymptotically converges to a maximum of a concave function.  SGD is a first-order optimisation method, which uses an unbiased estimate of the gradient of the objective function to decide which direction a step will be taken in. For any first-order optimisation algorithm with an appropriately chosen step size, the rate of convergence is generally $Q-$linear \citep{meng2020fast} for a function with a sufficiently smooth gradient. In the following subsections, we shall give more details on gradient descent, stochastic gradient descent and batch stochastic gradient descent algorithms.

\subsubsection{Gradient Descent (GD)}
The classical gradient descent algorithm was first proposed by \citet{cauchy1847}, but also independently by \citet{hadamard1908memoire}. It works on a function $ f \in \mathcal{C}^{1} $ (continuously differentiable) by taking steps in the direction of the negative gradient calculated at the current point $ \bm \theta $. If the objective is to be minimised, at a given point $ \bm \theta_t $, one step is defined as subtracting $ \alpha \nabla f(\bm \theta_t) $ from the current point. For a step size $ \alpha $ that is small enough, it holds that $ f(\bm \theta_t) \geq f(\bm \theta_{t+1}) $. This will lead to a monotonic sequence where $ f(\bm \theta) $ is decreasing, eventually reaching a minimum \citep{bottou2018optimization}. In our case, we are interested in maximizing the log-likelihood, i.e. $f(\bm \theta) = -\ln p(\mathbf{y}|\mathbf{x},\bm \theta)$, yet we shall continue this subsection with a general description of $f(\bm \theta)$.

A formal description of GD is provided in Algorithm~\ref{gd}, but many implementations with for example different stopping criteria exist. It is for instance possible to run until no improvements larger than a small number $\varepsilon$ are made, or for a set number of iterations.

\begin{algorithm}[H]
	\SetAlgoLined
	Gradient descent with convergence tolerance $ \varepsilon > 0 $ and step size $ \alpha $.\\
	Initialise a first solution $ \bm \theta_1 = \bm \theta_{\text{start}} $.\\
	\While{$ | \bm \theta_t - \bm \theta_{t-1}| > \varepsilon $}{
		Calculate the gradient at the current point, $ \nabla f(\bm \theta_t) $.\\
		Set $ \bm \theta_{t+1} = \bm \theta_t - \alpha_t \nabla f(\bm \theta_t) $.
	}
	\caption{Gradient Descent}
	\label{gd}
\end{algorithm}

\begin{figure}
	\centering
	\includegraphics[scale=0.15]{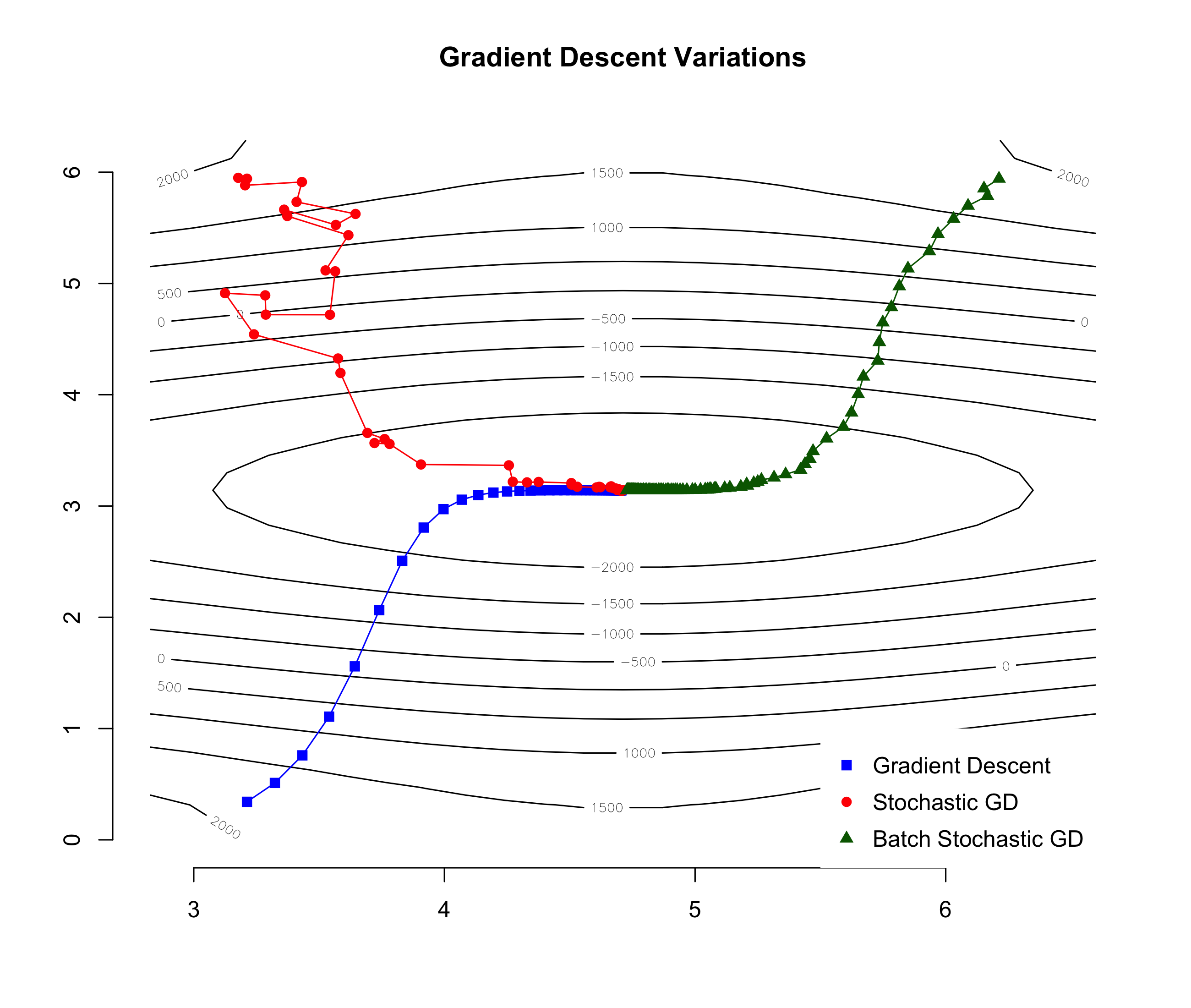}
	\caption{Demonstration of regular gradient descent (GD), stochastic gradient descent (SGD) and batch stochastic gradient descent (BSGD) applied to find a local minimum for $ f((a_i,b_i);\theta_1,\theta_2) = a_i \sin(\theta_1) + b_i \cos(\theta_2) $ where $ a_i \sim \text{N}(0.5,2^2) $ and $ b_i \sim \text{N}(-2,2^2) $. A total of 1,000 observations of $(a_i,b_i,f((a_i,b_i);\theta_1,\theta_2))$ were available, and BSGD used 50 per iteration.}
	\label{fig:gradient_descent}
\end{figure}

\subsubsection{Stochastic Gradient Descent (SGD)}
Originally proposed by \citet{robbinsmonro1951sgd}, \textit{Stochastic Gradient Descent (SGD)} can be considered as a modification of regular Gradient Descent. Let $ f(\bm \theta) = -\ln p(\mathbf{y}|\mathbf{x},\bm \theta)$ be a negative log likelihood function for a set of data $ (\bm y, \bm x) $. Assume there are $ n $ observations. SGD works by replacing the gradient at each iteration with an  approximation using only one data point, i.e. $ \widehat \nabla f(\bm \theta) = -\nabla_{\bm \theta} \ln p(y_i| x_i, \bm \theta) $ where $ i $ denotes one observation of the data chosen randomly at each iteration. The method also requires that the estimator of the gradient is unbiased, i.e. $E[ \widehat \nabla f(\bm \theta)]=\nabla f(\bm \theta)$. The latter is easily satisfied when the data are conditionally independent.

In contrast to using the full gradient, the unbiased approximation obtained will be very noisy and might for some steps cause the algorithm to make a move in an unfavourable direction. However, it is much less costly to compute, allowing for many more iterations to be performed inside the same computational budget. SGD has received a lot of recent attention in the machine learning community where it is the workhorse of neural networks among other techniques. An important reason for this is that it scales very well when the amount of data is large \citep{bottou2018optimization}. 

\subsubsection{Batch Stochastic Gradient Descent (BSGD)}
A middle ground between GD and SGD is \textit{Batch SGD (BSGD)}. Here the gradient is still estimated, but this is done using a subset or \textit{batch} of data at each iteration, with size $b$ chosen at the users' discretion. There exist variants that either go through the data in a fixed or random order. For the former, each batch consists of the $b$ observations located immediately after the previous (starting at the beginning when all batches have been visited). The other alternative is to randomly sample $b$ observations at each iteration. The iteration complexity when optimising the MLE is here $ \mathcal{O}(bm)$, where $m$ is the number of estimated parameters. Further, the convergence rate is generally between that of GD and SGD, owing to the more accurate (less noisy) approximation of the gradient \citep{bottou2018optimization}. 

In Figure \ref{fig:gradient_descent}, a demonstration of GD, SGD and BSGD is shown. Here, $b=50$ observations were used at each iteration for BSGD. The observations were obtained as a random sample from the total $1,000$ observations. We can see that gradient descent is able to converge to a local minimum in quite few iterations. The path from the starting point to the extremum is smooth and consistent. In the MLE setting, the iteration complexity is equal to that of evaluating the gradient $ \mathcal{O}(mn) $. Here, we can see that SGD is also able to converge to a local minimum in relatively few iterations. We should however note that the objective function in the figure is very smooth and does not pose a big challenge for any first- or second-order optimisation algorithm. The path for SGD is very jagged as the estimated gradient $ \widehat{\nabla} f(\bm \theta) $ does not always point towards the minimum. The iteration complexity is reduced to be just $ \mathcal{O}(m) $, the convergence rate is however an issue for less well-behaved functions than in the example. BSGD, in turn, gives a good trade-off in terms of smoothness and has  a complexity  $ \mathcal{O}(bm) $.
If applied to a GLM with the objective to maximise its likelihood, the SGD and BSGD converge to MLE. 

\subsubsection{Subsampling IRLS with BSGD}
The main drawback of GD, SGD and BSGD is that the required number of iterations may depend crucially on the quality of their starting points. If starting in a bad region of the parameter space, which is not unlikely for randomised starting points, it might take GD, SGD, and BSGD prohibitively long to converge. For GLM, the standard solution is to use second-order optimisation methods like IRLS. Combining second-order optimisation with subsampling, we propose a new algorithm for obtaining these starting points called Subsampling IRLS (S-IRLS). To obtain the MLE of a given model, S-IRLS is run for several iterations to generate good starting points for SGD or BSGD which in turn is used to find the final estimates. The overall idea is that S-IRLS is able to quickly provide a rough estimate of the parameters, and SGD (BSGD) is then run with the S-IRLS estimate as a starting point to practically and theoretically guarantee the convergence to the MLE. The resulting algorithm is abbreviated as S-IRLS-SGD in what follows. 

S-IRLS is presented generally in Algorithm~\ref{sub_irls_sgd}, for a detailed explanation see Appendix~\ref{sec:S-IRLS} of the article. In a small demonstration that follows, we shall see that while using S-IRLS is not critical, it may provide a boost in performance in the sense of computational time and accuracy of the estimates. In practice, we recommend doing this kind of initialisation within our approach to enhance additional robustness of its performance. The number of S-IRLS iterations is one of the tuning parameters $n_{\text{init}}$. It can be set rather small, i.e. below 100. To use random initialisation $n_{\text{init}} = 0 $ S-IRLS iterations for initialisation should be chosen. Then S-IRLS-SGD then collapses to either SGD or BSGD depending on the subsample size chosen.

Further, we provide an empirical evaluation of the performance of IRLS, BSGD, and S-IRLS-SGD. A set of data with one binary dependent variable, 15 independent variables and 1 million observations was generated. For details on the generative process, see Example 1 in Section~\ref{example1}. A Bernoulli regression with a g-prior from Example 1 in Section~\ref{example1} was addressed for this evaluation.
A full enumeration of all the 32,768 possible models was carried out using IRLS and the 128 models with the highest posterior mass were selected. These models were then re-evaluated using different optimisation algorithms of interest.  20 runs per model for each setting of each of the algorithms were performed. These algorithms and their tuning parameters are presented in Table \ref{table:optim}. This table also lists theoretical computational costs for the addressed methods and tuning parameters.
\begin{table}[]
\centering
\begin{tabular}{l|rrcccrr}
\hline
 & Iterations & \% of $n$ & $p+1$ & decay & $ \alpha_0 $ & Cost per step & Full cost\\
 \hline
IRLS & $5\times 10^0$ & $100\%$ & 16 & -  & - & $2.56\times 10^8$ & $1.28\times 10^9$ \\
BSGD 500 & $5\times 10^2$ & $0.10\%$ & 16 & 0.99995 & 0.20 & $1.6\times 10^4$ & $8\times 10^6$ \\
BSGD 1K & $1\times 10^3$ & $0.10\%$ & 16 & 0.99995 & 0.20 & $1.6\times 10^4$ & $1.6\times 10^7$ \\
BSGD 5K & $5\times 10^3$& $0.10\%$ & 16 & 0.99995 & 0.20 & $1.6\times 10^4$ & $8\times 10^7$ \\
BSGD 10K & $1\times 10^4$ & $0.10\%$ & 16 & 0.99995 & 0.20 & $1.6\times 10^4$ & $1.6\times 10^8$ \\
BSGD 20K & $2\times 10^4$ & $0.10\%$ & 16 & 0.99995 & 0.20 & $1.6\times 10^4$ & $3.2\times 10^8$ \\
\hline
S-IRLS & $7.5\times 10^1$ & $0.10\%$ & 16 & - & - & $2.56\times 10^5$ & $1.92\times 10^7$ \\
SGD & $5\times 10^2$ & $0.10\%$ & 16 & 0.99000 & 0.05 & $1.6\times 10^4$ & $8\times 10^6$ \\
S-IRLS-SGD & $5.75\times 10^2$ & $0.10\%$ & 16 & - & - & - & $2.72 \times 10^7$ \\
\hline
\end{tabular}
\vspace{5mm}
\caption{Hyper-parameters of the compared algorithms, computational costs per iteration and their expected computational costs. Here, a model with $p=15$ resulting in 16 parameters is assumed and the sample size $n$ is $10^6$ observations. Decay and $\alpha_0$ are the tuning parameters of exponential cooling for the sequence of step sizes $\alpha_t$ for BSGD optimisation routines \citep{bottou2007sgd}.}\label{table:optim}
\end{table}

\begin{figure}
    \centering

\includegraphics[scale=0.2]{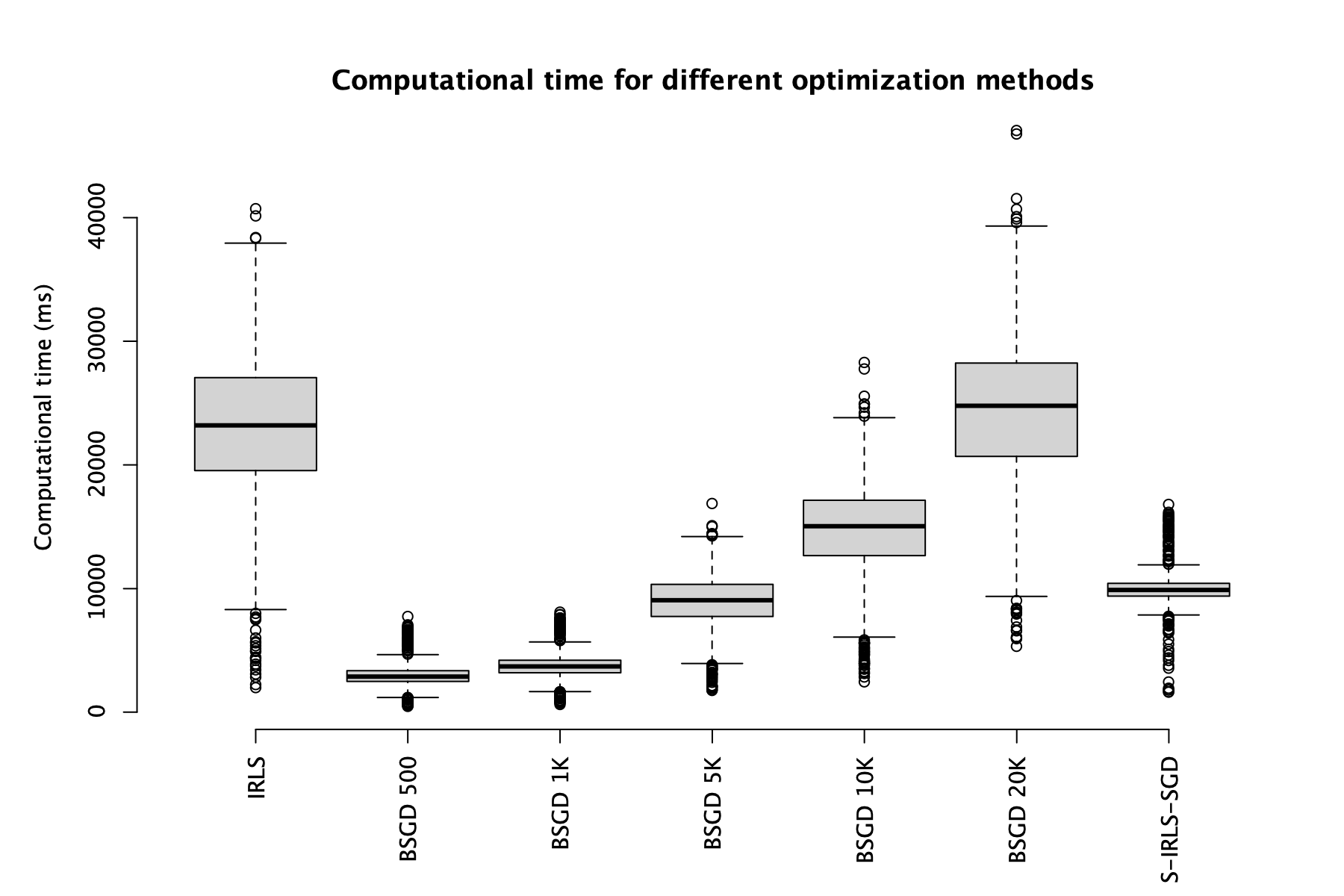}
    \caption{Boxplots of total computational time for different optimisation algorithms and tuning parameters from Table~\ref{table:optim}.}
    \label{fig:sgdstime}
\end{figure}

\begin{figure}[h]
    \centering
    \includegraphics[scale=0.2]{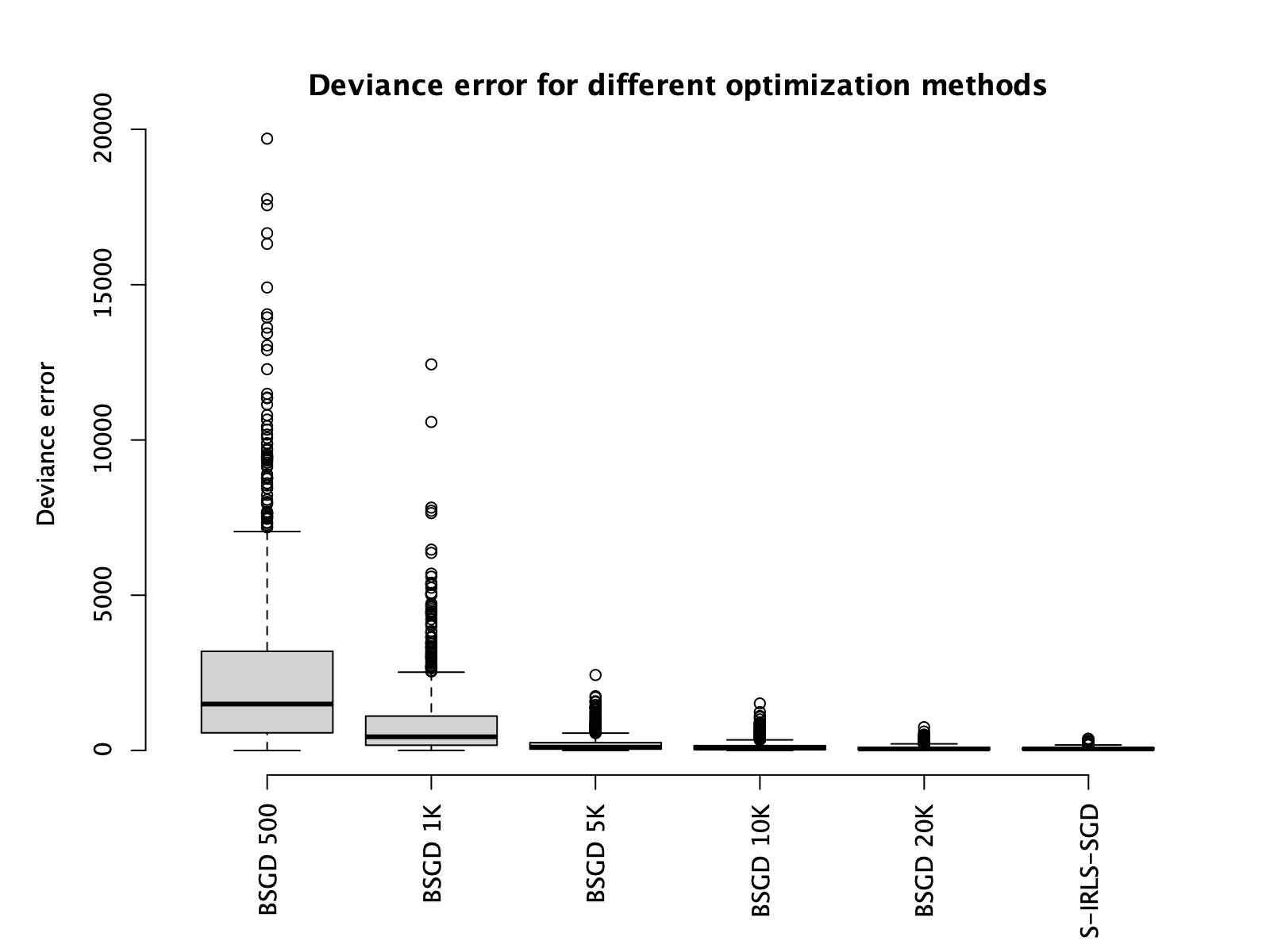}
    \caption{Boxplots of the errors of estimated deviances  for all different optimisation algorithms and tuning parameters from Table \ref{table:optim}. Four the last four ones see also Figure~\ref{fig:sgdserr2} in the Appendix~\ref{ap:A0} of the article, which gives a higher resolution.}\label{fig:sgdserr}
\end{figure}

In Figure~\ref{fig:sgdserr}, we present boxplots of the errors of estimated deviance. And in Figure \ref{fig:sgdstime}, the boxplots of computational times are given. As expected from the theoretical complexities of the algorithms from Table \ref{table:optim}, the addressed setting of S-IRLS-SGD initialisation gives a reasonable computational time compared to running BSGD to convergence or using the full IRLS. Furthermore, the errors of estimated deviance are the smallest for S-IRLS-SGD. For this reason, we shall focus on the S-IRLS-SGD variant of BSGD in the experiments to follow in Section \ref{section4}.

\begin{algorithm}[H]
\SetAlgoLined
\If {$n_{\text{init}} > 0$}
{Run $n_{\text{init}}$ iterations of S-IRLS to get a starting sub-optimal solution $ \hat{\bm \beta}_{\text{S}} $.}
\Else{
Randomise $ \hat{\beta}_{\text{S},i} \sim N(0,1)$ for all $i$.
}
Run SGD/BSGD using $  \hat{\bm \beta}_{\text{S}} $ as starting values to get the  solution $ \hat{\bm \beta}_{\text{BSGD}} $.\\
Calculate the deviance and likelihood using the full set of data and $ \hat{\bm \beta} $.
\caption{The suggested variant of BSGD for GLM.}
\label{sub_irls_sgd}
\end{algorithm}



\subsection{Combining MJMCMC/MCMC and BSGD} \label{sirlssgd_mjmcmc}

We have described some versions of the BSGD algorithm (SGD, S-IRLS-SGD) for finding the MLE of a GLM without requiring to use the full set of data at each iteration. This section will describe our proposed approach of combining the approximate MLE obtained from these algorithms with MJMCMC/MCMC for model selection.

Here, we begin by combining the Laplace approximation of the marginal likelihood from Equation~\eqref{margpost} with an approximate MLE. Denote the approximate MLE obtained from a version of BSGD or any other numerical optimisation algorithm as $ \bm{\tilde{\theta}}_{\mathfrak{m}} $. The approximation of the marginal likelihood is then
\begin{align}
    \ln \tilde p(\mathbf{y} | \mathfrak{m}) = \ln p(\mathbf{y} | \bm{\tilde{\theta}}_{\mathfrak{m}}) - \frac{|\bm{\tilde{\theta}}_{\mathfrak{m}}|}{2} \ln n. \label{mle_sirlssgd}
\end{align}
From the definition of the MLE
\begin{align}
    \bm{\hat{\theta}}_\mathfrak{m} = \underset{\bm \theta_\mathfrak{m} \in \bm \Theta_\mathfrak{m}}{\arg \max }\, p(\bm \theta_\mathfrak{m} | \mathbf{y},\mathfrak{m}),
\end{align}
combined with the monotonic property of the logarithm, we have that for any point $ \bm{\tilde \theta}_\mathfrak{m} \neq \bm{\hat{\theta}}_\mathfrak{m} $, the following holds
\begin{align}
    \ln p(\bm {\hat \theta_\mathfrak{m}} | \mathbf{y}, \mathfrak{m}) >  \ln p(\bm {\tilde \theta}_\mathfrak{m} | \mathbf{y}, \mathfrak{m}). \label{max_margpost}
\end{align}
We can therefore conclude that the approximation in Equation~\eqref{mle_sirlssgd} will be negatively biased every time a version of the BSGD algorithm is unable to fully converge to the exact MLE. This bias (if it is present) will under specific conditions diminish asymptotically when BSGD is combined with an MCMC algorithm. We shall address this in more detail in the next subsection.

Denote the marginal likelihood as 
\begin{align}\label{th_mlik}
p_L(\mathbf{y}|\mathfrak{m}) = \begin{cases} p(\mathbf{y}|\mathfrak{m}) \text{, if }  p(\mathbf{y}|\mathfrak{m}) \text{ is available}\\
\hat p(\mathbf{y}|\mathfrak{m}) \text{, otherwise}. 
\end{cases}
\end{align}

In this section, we aim at constructing an MCMC algorithm which has the following target:
\begin{align}\label{th_margpost}
    \pi(\mathfrak{m}) = \begin{cases}
p(\mathfrak{m}|\mathbf{y}) \text{ from Equation~\eqref{eq:fullpost}, if }p(\mathbf{y}|\mathfrak{m}) \text{ is available}\\
\hat p(\mathfrak{m}|\mathbf{y}) \text{ from Equation~\eqref{renormlaplace}, otherwise}.
\end{cases}
\end{align}
This target corresponds to either exact or Laplace approximated model probabilities~\eqref{renormlaplace}. In the latter case, the renormalised estimator is modified by replacing $p(\mathbf{y}|\mathfrak{m})$ by
$\hat p(\mathbf{y}|\mathfrak{m})$ in~\eqref{renorm}.
As shown in Equation~\eqref{max_margpost}, any point $ \bm{\tilde \theta}_\mathfrak{m} \neq \bm{\hat{\theta}}_\mathfrak{m} $ will result in a Laplace approximation of the marginal likelihood that is negatively biased in comparison to the one evaluated at the MLE. The size of this bias will be a function of how well the optimisation algorithm was able to converge. In this section, we present the algorithm describing the proposed approach together with some theory on how to combine BSGD (SGD, S-IRLS-SGD) with MCMC or MJMCMC. We also give some practical implications of this theory.

\begin{algorithm}[H]
	\SetAlgoLined
	Initialise a model $ \mathfrak{m}$ and assume for all models $ \mathfrak{m}'$ that $\hat{p}^{\text{opt,0}}(\mathbf{y} | \mathfrak{m}') = \epsilon, \epsilon \in (0, \min_{\bm{\tilde{\theta}_\mathfrak{m^{\star}}}, \mathfrak{m^{\star}} \in \mathcal{M}}  \hat{p}(\mathbf{y}|\mathfrak{m^{\star}})]$ with $\hat{p}(\mathbf{y}|\mathfrak{m^{\star}})$ being a function of $\bm{\tilde{\theta}_\mathfrak{m^{\star}}}$ from \eqref{laplace}.\\
	\For{$ s \in \{1,...,t\}$}{
		Propose a model $\mathfrak{m}^{\star}$ using MCMC proposal kernel from a mixture $q_r(\mathfrak{m}^{\star}|\mathfrak{m}), r \in R$.\\
	    Obtain $\hat{\theta}_{\text{BSGD},\mathfrak{m}^{\star}}$ by means of Algorithm~\ref{sub_irls_sgd}.\\
	    \If{$\text{Bernoulli}(p_{\text{rand}})=1$}
{Make a small randomisation $\hat{\theta}_{s,\mathfrak{m}^{\star}} \sim N(\hat{\theta}_{\text{BSGD},\mathfrak{m}^{\star}},\sigma^2_{\text{rand}})$ to get the final $ \hat{\theta}_{s,\mathfrak{m}^{\star}}$.}
\Else{Set $\hat{\theta}_{s,\mathfrak{m}^{\star}} = \hat{\theta}_{\text{BSGD},\mathfrak{m}^{\star}}$}
	    Compute $\hat p^{s}(\mathbf{y} | \mathfrak{m}^{\star})$ using  $\hat{\theta}_{s,\mathfrak{m}^{\star}}$  and the whole data $\mathbf{y}$ according to \eqref{mle_sirlssgd}.\\
	    Set  \begin{align}\label{xyzzyx}
	        \hat p^{\text{opt,s}}(\mathbf{y} | \mathfrak{m}^{\star}) = \max \left(\hat p^{\text{opt,s-1}}(\mathbf{y} | \mathfrak{m}^{\star}), \hat p^{s}(\mathbf{y} | \mathfrak{m}^{\star})\right).
	    \end{align}\\
	    Set $\hat p^{\text{opt,s}}(\mathbf{y} | \mathfrak{m}') = \hat p^{\text{opt,s-1}}(\mathbf{y} | \mathfrak{m}')$ for all $\mathfrak{m}'\neq\mathfrak{m}^{\star}$.\\
	    Set the next state of the chain  to\\
		\begin{align}
			\mathfrak{m}' = \begin{cases}
				\mathfrak{m}^{\star} & \text{with probability } \alpha = r^{\star}_{mh}(\mathfrak{m}, \mathfrak{m}^{\star};s);\\
				\mathfrak{m} & \text{otherwise,}
			\end{cases}
		\end{align}
	 	where
		\begin{align}
			r^{\star}_{mh}(\mathfrak{m}, \mathfrak{m}^{\star};s) = \min \left\lbrace 1, \frac{\hat p^{\text{opt,s}}(\mathbf{y} | \mathfrak{m}^{\star})p(\mathfrak{m}^{\star}) q (\mathfrak{m} | \mathfrak{m}^{\star})}{\hat p^{\text{opt,s}}(\mathbf{y} | \mathfrak{m})p(\mathfrak{m}) q(\mathfrak{m}^{\star} | \mathfrak{m})} \right\rbrace.
		\end{align}
	    
	}
	\caption{MCMC with BSGD.}
	\label{sub_mcmc_sgd}
\end{algorithm}

 The general algorithm for combining MCMC for model exploration with BSGD for computing the MLIKs is presented in Algorithm~\ref{sub_mcmc_sgd}. To achieve the theoretical convergence properties that will be proven in Theorem~\ref{theorem3}, we allow for a small independent normal randomisation with $\sigma^2_{\text{rand}}\ll 1$ with probability $p_{\text{rand}}$ around the estimated by MLE before calculating the marginal likelihood. Note that in our implementation, the MJMCMC version of MCMC is combined with the S-IRLS-SGD version of BSGD, which guarantees by design that a in mixture of proposal kernels $q_r(\mathfrak{m}^\star|\mathfrak{m}), r \in R$ and $\exists r'$ such that $q_{r'}(\mathfrak{m}^\star|\mathfrak{m})>0, \forall \mathfrak{m},\mathfrak{m}^\star \in \mathcal{M}$ required for Lemma~\ref{lemma1}.
 

\begin{lemma} \label{lemma1} Let a finite and countable model class $\mathcal{M}$ with $\pi(\mathfrak{m}) \in (0,1), \forall \mathfrak{m} \in \mathcal{M}$, $\sum_{\mathfrak{m}\in\mathcal{M}}\pi(\mathfrak{m}) = 1$,  and  $p(\bm \theta_\mathfrak{m}|\mathfrak{m})p(\mathfrak{m}) > 0, \forall \mathfrak{m} \in \mathcal{M}, \theta_\mathfrak{m} \in \bm \Theta_\mathfrak{m}$ with $\mathcal{M}$ being a space of states of the MCMC Algorithm~\ref{sub_mcmc_sgd} with a mixture of proposal kernels $q_r(\mathfrak{m}^\star|\mathfrak{m}), r \in R$ and $\exists r'$ such that $q_{r'}(\mathfrak{m}^\star|\mathfrak{m})>0, \forall \mathfrak{m},\mathfrak{m}^\star \in \mathcal{M}$. Then under the subsampling scheme from Algorithm~\ref{sub_mcmc_sgd}, the Markov chain induced by the MCMC algorithm is irreducible and recurrent.
\begin{proof}
Define $p^0=\min_{\mathfrak{m} \in \mathcal{M}}p^{opt,0}(\bm y|\mathfrak{m})$ and $p^{\infty}=\max_{\mathfrak{m}' \in \mathcal{M}}\hat p^{opt,\infty}(\bm y|\mathfrak{m})$. Then $0<p_0\le \hat p^{opt,0}(\bm y|\mathfrak{m})\le p^{opt,0}(\bm y|\mathfrak{m})\le p^{\infty}<\infty $ which gives
\begin{align*}
p^t(\mathfrak{m}|\bm y)=\frac{p(\mathfrak{m})p^{opt,0}(\bm y|\mathfrak{m})}{\sum_{\mathfrak{m}}'p(\mathfrak{m}')p^{opt,t}(\bm y|\mathfrak{m}')}
\ge \frac{p(\mathfrak{m})p^0}{\sum_{\mathfrak{m}}'p(\mathfrak{m}')p^{\infty}}=\frac{p^0}{p^{\infty}}p(\mathfrak{m}),     
\end{align*}
showing that there is a lower bound for the model probabilities and thereby a lower bound for the transition probabilities in the Markov chain. This implies irreducibility. Since the space of states is discrete and finite, the Markov chain will also be recurrent.

\end{proof}
\end{lemma}
\begin{lemma} \label{lemma2}
Let for every visited by Algorithm~\ref{sub_mcmc_sgd} model $\mathfrak{m} \in \mathcal{M}$, there exists a sequence of parameters that converges (a) almost surely $\bm \hat{\theta}_{t,\mathfrak{m}} \xrightarrow{a.s.} \bm \theta_{\text{MLE},\mathfrak{m}}$ or (b) in probability $\bm \hat{\theta}_{t,\mathfrak{m}} \xrightarrow{p} \bm \theta_{\text{MLE},\mathfrak{m}}$. Then, for the estimates of marginal likelihoods $\hat p^{\text{opt,t}}(\mathbf{y} | \mathfrak{m})$ defined in Equation~\eqref{xyzzyx} it holds for (a) that  $p^{\text{opt,t}}(\mathbf{y} | \mathfrak{m})\xrightarrow{a.s.} p_L(\mathbf{y}| \mathfrak{m})$ and for (b) that $p^{\text{opt,t}}(\mathbf{y} | \mathfrak{m})\xrightarrow{p} p_L(\mathbf{y}| \mathfrak{m})$. And for the target model probabilities ${\pi(\mathfrak{m})}$ it holds for (a) that $\hat p_{RM}^{(t)}( \mathfrak{m}|\mathbf{y})\xrightarrow{a.s.} {\pi(\mathfrak{m})}$ as $t\rightarrow\infty$ and for (b) that $\hat p_{RM}^{(t)}( \mathfrak{m}|\mathbf{y})\xrightarrow{p} {\pi(\mathfrak{m})}$.

\begin{proof}
By Lemma~\ref{lemma1} every model will eventually be visited as $t\rightarrow\infty$. Also, by regularity conditions, there exists a sequence of parameters that converges (a) almost surely $\bm \hat{\theta}_{t,\mathfrak{m}} \xrightarrow{a.s.} \bm \theta_{\text{MLE},\mathfrak{m}}$ or (b) in probability $\bm \hat{\theta}_{t,\mathfrak{m}} \xrightarrow{p} \bm \theta_{\text{MLE},\mathfrak{m}}$.
Consequently, by the continuous mapping theorem \citep{mann1943stochastic}, as $t\rightarrow\infty$ we have for (a) that $\hat p^{\text{opt,t}}(\mathbf{y}|\mathfrak{m})\xrightarrow{a.s.}{p(\mathbf{y}|\mathfrak{m})}$  and for (b) that $\hat p^{\text{opt,t}}(\mathbf{y}|\mathfrak{m})\xrightarrow{p}{p(\mathbf{y}|\mathfrak{m})}$. Combining this with the fact that $\mathcal{M}$ is finite and countable and following Bayes theorem and properties of convergence of the sums and products of sequences converging with probability one as $t\rightarrow\infty$ we get for (a) that $\hat p_{RM}^{(t)}( \mathfrak{m}|\mathbf{y})\xrightarrow{a.s.}{\pi(\mathfrak{m})}$ and for (b) that $\hat p_{RM}^{(t)}( \mathfrak{m}|\mathbf{y})\xrightarrow{p}{\pi(\mathfrak{m})}$, where $\hat p_{RM}^{(t)}(\mathfrak{m}|\mathbf{y})$ are obtained through Equation~\eqref{renorm} with  $\hat p^{\text{opt,t}}(\mathbf{y} | \mathfrak{m})$ plugged in for  $ p(\mathbf{y} | \mathfrak{m})$.
\end{proof}

\end{lemma}
%
%

We further provide three theorems under various regularity conditions. Consider first some general Conditions~\ref{cond1}-\ref{cond5}.

\begin{condition} \label{cond1}
    $\ln p(\mathbf{y} |\bm \theta_\mathfrak{m} ,\mathfrak{m})$ is Lipschitz continuous with Lipschitz constant $G > 0$ and its gradient is Lipschitz continuous with Lipschitz constant $L > 0$, i.e. $|\ln p(\mathbf{y} |\bm \theta_\mathfrak{m} ,\mathfrak{m})-\ln p(\mathbf{y} |\bm \theta'_\mathfrak{m} ,\mathfrak{m})|\leq G||\theta_\mathfrak{m}-\theta'_\mathfrak{m}||$ and $||\nabla\ln p(\mathbf{y} |\bm \theta_\mathfrak{m} ,\mathfrak{m})-\nabla\ln p(\mathbf{y} |\bm \theta'_\mathfrak{m} ,\mathfrak{m})||\leq L||\theta_\mathfrak{m}-\theta'_\mathfrak{m}||$.
    \end{condition}
    \begin{condition} \label{cond2}
    $\ln p(\mathbf{y} |\bm \theta_\mathfrak{m} ,\mathfrak{m})$ is strictly concave, i.e. $\ln p( \mathbf{y} | (1-a)\bm \theta_\mathfrak{m} + a\bm \theta'_\mathfrak{m}, \mathfrak{m})>(1-a)\ln p(\mathbf{y} |\bm \theta_\mathfrak{m} ,\mathfrak{m})+a\ln p(\mathbf{y} |\bm \theta'_\mathfrak{m} ,\mathfrak{m}), \forall a \in [0,1]$.
    \end{condition}
\begin{condition} \label{cond3}
$ - \ln  p(\mathbf{y} |\bm \theta_\mathfrak{m} ,\mathfrak{m})$ is coercive, i.e. $- \ln  p(\mathbf{y} |\bm \theta_\mathfrak{m} ,\mathfrak{m}) \rightarrow \infty$ as $||\bm \theta_\mathfrak{m}||\rightarrow \infty$.
\end{condition}
\begin{condition} \label{cond4}
$-\ln  p(\mathbf{y} |\bm \theta_\mathfrak{m} ,\mathfrak{m})$ is asymptotically flat. i.e. $\lim \inf_{||\bm \theta_\mathfrak{m}||\rightarrow\infty} ||\nabla\ln p^{-1}(\mathbf{y} |\bm \theta_\mathfrak{m} ,\mathfrak{m})||>0$.
\end{condition}

\begin{condition} \label{cond5} 
$\nabla\ln  p(\mathbf{y} |\bm \theta_\mathfrak{m} ,\mathfrak{m})$ can be accessed via a stochastic first order oracle, i.e. we have $V(\bm \theta_\mathfrak{m};w) = \nabla\ln  p(\mathbf{y} |\bm \theta_\mathfrak{m} ,\mathfrak{m}) + Z(\bm \theta_\mathfrak{m};w)$, where the error term $Z(\bm \theta_\mathfrak{m};w)$ has a zero mean $E\{Z(\bm \theta_\mathfrak{m};w)\} = 0$ and finite moments $E\{||Z(\bm \theta_\mathfrak{m};w)||^q\} \leq \sigma^q$ for some $q\geq 2$ and $\sigma\geq 0$.
\end{condition}

Also, consider the case where BSGD is run with an upper bound $U$ on the maximal number of allowed iterations, which will cause it to not fully converge to the MLE at least some times for some of the models in the model space.  At the same time, assume that BSGD starts at the values of parameters of the model it finished last time. Then, the following Theorem gives a general result that the algorithm will still converge to the true model posteriors.

\begin{theorem} \label{theorem2}
Let for every model $\mathfrak{m} \in \mathcal{M}$ visited by Algorithm~\ref{sub_mcmc_sgd}, the BSGD optimisation algorithm is run for up to $U$ iterations with a step-size sequence of the form $\alpha_\tau = \Theta(1/\tau^l)$ for some $l\in(2/(q+2),1]$. Let also Conditions~\ref{cond1}-\ref{cond5} hold $\forall \mathfrak{m} \in \mathcal{M}$. Let for every revisit of a model, BSGD start at the values of parameters of the model and with the same tuning parameters of BSGD as it finished during the previous visit of that model. Then, for the estimates of marginal likelihoods $\hat p^{\text{opt,t}}(\mathbf{y} | \mathfrak{m})$ defined in Equation~\eqref{xyzzyx} it holds that   $p^{\text{opt,t}}(\mathbf{y} | \mathfrak{m})\xrightarrow{a.s.} p_L(\mathbf{y}| \mathfrak{m})$ and for the target model probabilities ${\pi(\mathfrak{m})}$ it holds that $\hat p_{RM}^{(t)}( \mathfrak{m}|\mathbf{y})\xrightarrow{a.s.} {\pi(\mathfrak{m})}$  as $t\rightarrow\infty$.

\begin{proof}

By Lemma~\ref{lemma1} every model will eventually be visited infinitely often as $t \rightarrow \infty$. Given that Conditions~\ref{cond1}-\ref{cond5} hold for all $\mathfrak{m} \in \mathcal{M}$, the regularity conditions of Theorem 2 from \citet{mertikopoulos2020almost} also hold and hence as $t\rightarrow\infty$ we have $\bm \hat{\theta}_{t,\mathfrak{m}} \xrightarrow{a.s.} \bm \theta_{\text{MLE},\mathfrak{m}}$ where $\bm \theta_{\text{MLE},\mathfrak{m}}$ is the unique maximal likelihood estimate of parameters for model $\mathfrak{m}$ as the log-likelihood function is strictly concave. Consequently, the regularity conditions of Lemma~\ref{lemma2} hold and thus for $\hat p^{\text{opt,t}}(\mathbf{y} | \mathfrak{m})$ defined in Equation~\eqref{xyzzyx} it holds that   $p^{\text{opt,t}}(\mathbf{y} | \mathfrak{m})\xrightarrow{a.s.} p_L(\mathbf{y}| \mathfrak{m})$ and  $\hat p_{RM}^{(t)}( \mathfrak{m}|\mathbf{y})\xrightarrow{a.s.} {\pi(\mathfrak{m})}$ as $t\rightarrow\infty$.
\end{proof}
\end{theorem}

\begin{remark}
Following Theorem~\ref{theorem2}, the bias will reduce to zero with probability one if we allow MCMC (MJMCMC) to run until convergence and we reestimate each revisited model from a point where we finished last time by starting with the tuning parameters of BSGD with which we finished last time for that model.
\end{remark}

\begin{corollary} Under the regularity conditions of Theorem \ref{theorem2},  assume that the adaptive transition kernel $ q_r(\mathfrak{m}^{\star}|\mathfrak{m})	r^{\star}_{mh}(\mathfrak{m}, \mathfrak{m}^{\star};t), r \in R$ satisfies assumption A3 from \citet{fort2011convergence} and that for this adaptive transition kernel and for $\pi^t(\mathfrak{m})$ (i)-(iii) from Theorem 2.11 from \citet{fort2011convergence} hold, then  Monte Carlo estimates of model probabilities based on the frequencies of a MCMC (MJMCMC) algorithm will almost surely converge to the true posteriors $\hat{p}_{MC}^{(t)}(\mathfrak{m}|\mathbf{y})\xrightarrow{a.s.} {\pi(\mathfrak{m})}$ as $t\rightarrow\infty$. \label{corollary1}
\begin{proof} 
$\mathcal{M}$ is a finite discrete countable space of models, which is completely metrizable and separable. Hence, it is a Polish space. Consequently, the proof follows directly from Theorem 2.11 in \citet{fort2011convergence}.


\end{proof}
\end{corollary}

Now, we shall show that for both concave and non-concave likelihoods, if BSGD starts every new visit of a model at any point in parameter space (either randomly or using any kind of a rule of optimisation combined with randomisation) and if we allow using the randomisation at the end of BSGD optimisation, then the renormalised estimates produced by Algorithm~\ref{sub_mcmc_sgd} will converge in probability.

\begin{theorem} \label{theorem3}
Let the regularity conditions of Lemma~\ref{lemma1} hold. Also, allow with probability $p_{\text{rand}}>0$ for a randomisation around the approximated MLE $\bm{\hat \theta}_{s,\mathfrak{m}}$ from Equation~\eqref{mle_sirlssgd} from a probability distribution $ u({\bm \theta_{\mathfrak{m}}}) $ covering the support for the full parameter space $\bm {\Theta}_{\mathfrak{m}}$. Then, for the estimates of marginal likelihoods $\hat p^{\text{opt,t}}(\mathbf{y} | \mathfrak{m})$ defined in Equation~\eqref{xyzzyx} it holds that   $p^{\text{opt,t}}(\mathbf{y} | \mathfrak{m})\xrightarrow{p} p_L(\mathbf{y}| \mathfrak{m})$ and for the target model probabilities ${\pi(\mathfrak{m})}$ it holds that $\hat p_{RM}^{(t)}( \mathfrak{m}|\mathbf{y})\xrightarrow{p} {\pi(\mathfrak{m})}$  as $t\rightarrow\infty$.

\begin{proof}
By Lemma~\ref{lemma1}, each model $\mathfrak{m}\in\mathcal{M}$ will be visited infinitely often as $t\rightarrow\infty$. Then, by the design of $ u({\bm \theta_{\mathfrak{m}}}) $, the regularity conditions of the theorem from \citet{matyas1965random} hold (for a later reference of the same result see also \citet{solis1981minimization}) and as $t\rightarrow\infty$ we have $\bm \hat{\theta}_{t,\mathfrak{m}} \xrightarrow{p} \bm \theta_{\text{MLE},\mathfrak{m}}$ where $\bm \theta_{\text{MLE},\mathfrak{m}}$ is the maximal likelihood estimate of parameters for model $\mathfrak{m}$. Consequently, the regularity conditions of Lemma~\ref{lemma2} hold and thus for $\hat p^{\text{opt,t}}(\mathbf{y} | \mathfrak{m})$ defined in Equation~\eqref{xyzzyx} it holds that   $p^{\text{opt,t}}(\mathbf{y} | \mathfrak{m})\xrightarrow{p} p_L(\mathbf{y}| \mathfrak{m})$ and  $\hat p_{RM}^{(t)}( \mathfrak{m}|\mathbf{y})\xrightarrow{p} {\pi(\mathfrak{m})}$ as $t\rightarrow\infty$.
\end{proof}
\end{theorem}

\begin{remark}
Following Theorem~\ref{theorem3}, the bias will reduce to zero in probability if we allow MCMC (MJMCMC) to run until convergence.
\end{remark}
\begin{remark}
Theorem~\ref{theorem3} will hold both when we reestimate each revisited model from a random starting point or the best previously obtained point.
\end{remark}
\begin{corollary} Under the regularity conditions of Theorem \ref{theorem3}, assume that the adaptive transition kernel $ q_r(\mathfrak{m}^{\star}|\mathfrak{m})	r^{\star}_{mh}(\mathfrak{m}, \mathfrak{m}^{\star};t), r \in R$ satisfies assumption A3 from \citet{fort2011convergence} and that for this adaptive transition kernel and for $\pi^t(\mathfrak{m})$ (i)-(iii) from Theorem 2.11 from \citet{fort2011convergence} hold, then  Monte Carlo estimates of model probabilities based on the frequencies of a MCMC (MJMCMC) algorithm will almost surely converge to the true posteriors $\hat{p}_{MC}^{(t)}(\mathfrak{m}|\mathbf{y})\xrightarrow{a.s.} {\pi(\mathfrak{m})}$ as $t\rightarrow\infty$. \label{corollary2}
\begin{proof} 
Same as for Corollary~\ref{corollary1}.

\end{proof}
\end{corollary}
\begin{remark}
Following Corollaries~\ref{corollary1} and~\ref{corollary2}, the bias of MC estimates will reduce to zero almost surely if we allow MCMC (MJMCMC) to run until convergence.
\end{remark}
\begin{remark}
We were not able to formally check if the additional regularity conditions of Corollaries~\ref{corollary1} and~\ref{corollary2} from \citet{fort2011convergence} fully hold for Algorithm~\ref{sub_mcmc_sgd}. However, in Section~\ref{section4}, we see that empirically $\hat{p}_{MC}^{(t)}(\mathfrak{m}|\mathbf{y})$ converge to the underlying $\pi(\mathfrak{m})$. 
\end{remark}

In practice, Theorems~\ref{theorem2} and~\ref{theorem3} can be applied when using a fixed computational budget and will allow minimising the perturbation of the posterior induced by using an approximated MLE. In Theorem~\ref{theorem3}, for the parameters having the support on the real line, the randomisation around the approximated MLE can be for example $\text{N}(0, \sigma^2_{\text{rand}})$, where $\sigma^2_{\text{rand}}$ is small. Also, $p_{\text{rand}}$  can be selected to be very small. This will not affect the actual performance but still provide theoretical convergence. The conditions of Theorem~\ref{theorem3} are easily satisfied when using Algorithm~\ref{sub_mcmc_sgd}. 

\section{Experiments}\label{section4}

In this section, we first evaluate how S-IRLS-SGD based on different sub-sample sizes with multiple revisits of each model works for estimating the marginal posterior inclusion probabilities under full enumeration. Then, we show how  S-IRLS-SGD  works in combination with MJMCMC. 

\subsection{Benchmarks and tests of S-IRLS-SGD}
To evaluate the performance of the S-IRLS-SGD algorithm, it is directly compared to regular IRLS. The parameter of interest that is measured is the \textit{posterior marginal probability (PMP)} of including the independent variables $ p(\gamma_j = 1 | \mathbf{y}) $. S-IRLS-SGD was set to run 20 S-IRLS iterations for initialisation followed by 250 BSGD iterations for Gaussian models and 75 S-IRLS and 500 BSGD iterations for logistic models for all the experiments.

\subsubsection{Example 1. Simulated data} \label{example1}
To test the performance of S-IRLS-SGD for computing the marginal likelihoods under both full enumeration and MJMCMC, a simulated dataset was generated. It consisted of one dependent variable $ \mathbf{y} $, and 15 independent variables $ \bm x $. We explored the cases with $n$ equal to $10,000$ and $100,000$ observations. By combining the independent variables in all possible ways we can create $ 2^{15}=32,768 $ different linear models.

For our models, Zellner's g-prior \citep{zellner1986bayesian} was used with the parameter $$ g = \frac{100}{\sqrt{n/100}}. $$ This heavily penalises models that include many parameters, which in turn empirically ensures that a single model will not contain most of the posterior mass, avoiding PMPs to collapse to either 0 or 1. The latter is important to ensure that the true PMPs are challenging to estimate, making it is easy to examine how well the algorithm is performing.

\begin{figure}[!h]
	\centering
	\includegraphics[scale=0.175]{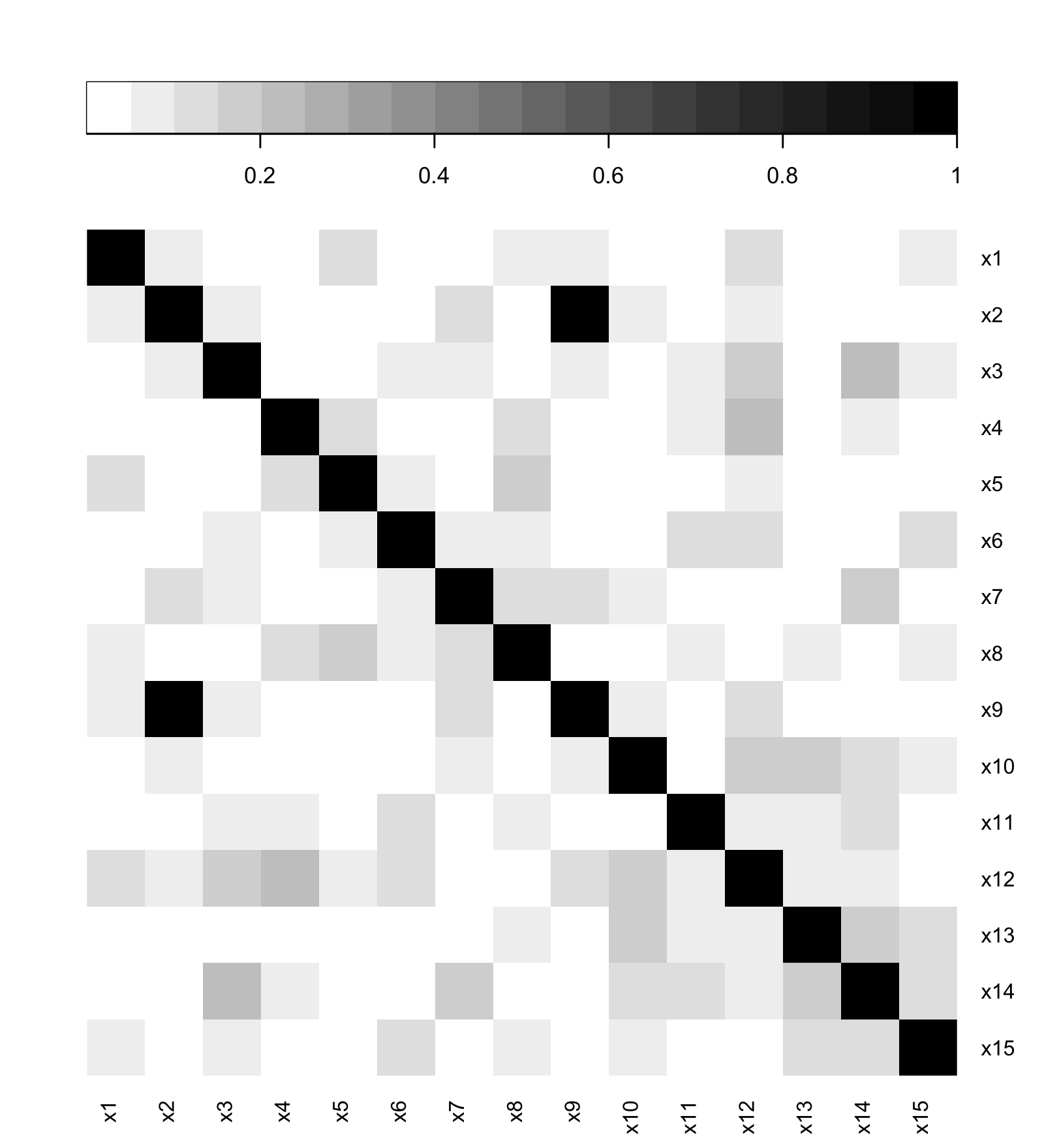}
	\caption{Absolute correlation between the exogenous $ x $ variables for the data used in Example~\hyperref[example1]{1}. The two variables $ x_2, x_9 $ are highly correlated, with the remaining variables having low to moderate correlation.}
	\label{fig:ex3_cor}
\end{figure}

The data was generated in a way to be comparable to the examples in  \citet{clyde2011bas,hubin2016efficient}. This means that we used the same covariance matrix for the independent variables $ \bm x $, and the same vector of parameters $ \bm \beta = (0.48, 8.72, 1.76, 1.87, 0, 0, 0, 0, 4, 0, 0, 0, 0, 0, 0) $ when generating the dependent variable $ \mathbf{y} $. The covariance structure of the data is illustrated in Figure~\ref{fig:ex3_cor}. As we performed tests with different sample sizes, adjustments were made to the $ \bm \beta $ vector to correct for this. For a given observation count $ n $ we have used
\begin{align}
	\bm \beta^{\star} = \frac{\bm \beta}{\sqrt{n/100}}, \label{signal}
\end{align}
which will give approximately the same signal strength in the data as $ \bm \beta $ did for 100 observations in \citet{hubin2020logic}. The dependent variables $ \mathbf{y} $ for $ n $ observations was generated according to the following distribution:
\begin{align}
	{y_i} \sim \text{N}(\bm x_i^{\text{T}} \bm \beta^{\star}, 1).
\end{align}
To be able to use the same data for not only Gaussian models, but also logistic ones, additional dependent variables $ { y_i}^{\star}$ were generated following the distribution, that in our example empirically does not lead to a perfect separation between the zeros and ones
\begin{align}
{ y_i}^{\star} \sim \text{Bernoulli}\left(\frac{1}{1+ \exp \left( - ({y_i} - \bar{{y}_i}) \right)} \right).
\end{align}
For this example, we performed full enumeration of all the 32,768 possible models using regular IRLS and S-IRLS-SGD with subsample sizes of 20\%, 10\%, 5\%, 1\%, 0.75\%, 0.5\%, 0.25\%, 0.1\% and 0.05\% on both $10,000$ and $100,000$ observations and for both Gaussian and logistic models. Each experiment was repeated 20 times in order to create confidence bands for the estimates.

\begin{figure}
	\centering
	\includegraphics[scale=0.18]{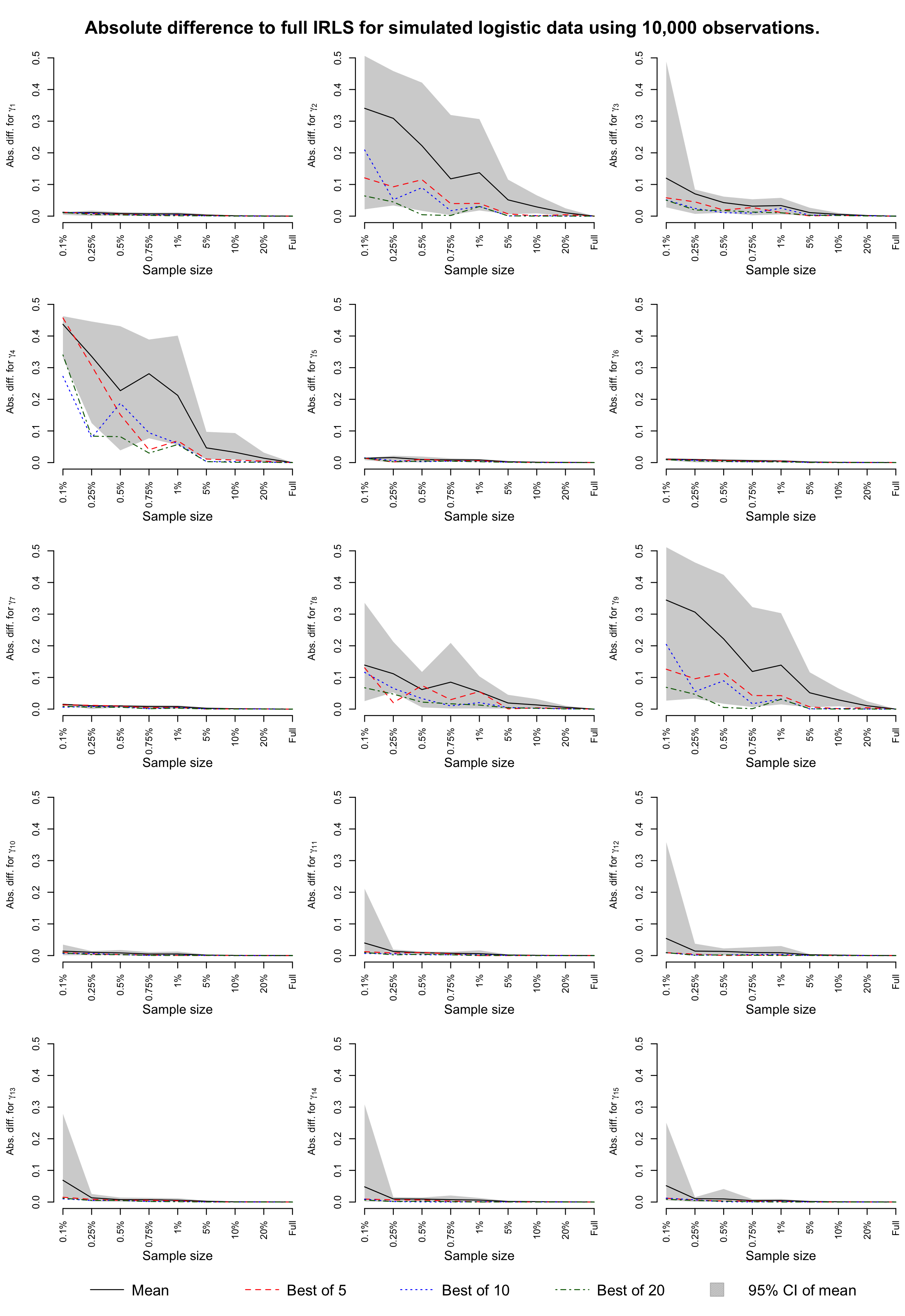}
	\caption{The absolute difference for marginal posterior probabilities for full enumeration of 32,768 logistic models using $10,000$ observations from the set of data in Example~\hyperref[example1]{1}. 20 runs were performed and the black line shows the mean absolute difference, with grey 90\% confidence intervals. The dashed and dotted lines show the absolute difference obtained by using the best likelihood estimate for each model that occurred during 5, 10 and 20 random runs.}
	\label{fig:full_10Kl}
\end{figure}

For each subsample size in S-IRLS-SGD, the absolute difference to the true value was calculated for the PMP of each $ \gamma_j $. The same was also done using the best estimate for each model from 5, 10 and 20 runs of the algorithm. The mean of the 20 runs for $10,000$ observations of logistic models, with 95\% confidence intervals is presented in Figure~\ref{fig:full_10Kl}. Also, we illustrate lines for the estimates created from the 5, 10, and 20 runs by selecting the maximum estimate for each model across these runs. The most apparent tendency is that for a larger subsample used in S-IRLS-SGD the estimates are closer to the true values. We also notice that when selecting the best estimate per model using a higher number of runs, the absolute difference to the true values gradually decreases. 

Additional plots for $100,000$ observations with logistic models, as well as $10,000$, and $100,000$ observations with Gaussian models are available in Figures~\ref{fig:full_10Kg}, \ref{fig:full_100Kg} and \ref{fig:full_100Kl} in Appendix~\ref{add_results}. In these figures, we see similar trends.

\subsection{Combining MJMCMC and S-IRLS-SGD}

In Example~\hyperref[example2]{2}, we apply the strategy with restarts of S-IRLS-SGD for every visit of a model inside Algorithm \ref{sub_mcmc_sgd}, corresponding to Theorem~\ref{theorem3}. We show empirically that this approach in the context of Bayesian variable selection and MJMCMC achieves a \textit{better} (in the sense of RMSE from the true posteriors) estimation than the full enumeration of the same model set and the same MLE approximation method on a given sub-sample size.

In these experiments, using the renormalised estimator \eqref{inclusion2}, we obtain an estimate that we can compare with the true values. Following \citet{clyde2011bas} and \citet{hubin2018mjmcmc}, we use \textit{Root Mean Square Error (RMSE)} which for $ K $ independent runs of the MJMCMC algorithm can be calculated using
\begin{align}
	\text{RMSE}(\gamma_j) = \sqrt{\frac{\sum_{i=1}^{K}\left[\hat{p}_K(\gamma_j=1 | \mathbf{y})-{p}(\gamma_j=1 | \mathbf{y})\right]^2}{K}}. \label{eq:rmse}
\end{align}

\subsubsection{Example 2: Simulated data (MJMCMC)} \label{example2}

Using the data from Example~\hyperref[example1]{1}, MJMCMC was run using both regular IRLS and S-IRLS-SGD with a subsample size of 5\%, 1\%, 0.75\%, 0.5\% and 0.25\%. The algorithm was run 20 times for each subsample size and each run was allowed to perform 33,000 iterations with no burn-in allowed. Using both the MC \eqref{inclusion1} and RM \eqref{inclusion2} estimators, the marginal inclusion probabilities for the 15 covariates were estimated. These estimates were compared to the true values by calculation of the RMSE as defined in Equation~\eqref{eq:rmse}. S-IRLS-SGD was set to run 20 S-IRLS and 250 BSGD iterations for Gaussian models and 75 S-IRLS  and 500 BSGD iterations for logistic data for all the experiments, where  S-IRLS is used for initialisation.
\begin{figure}[h]
	\centering
	\includegraphics[scale=0.2]{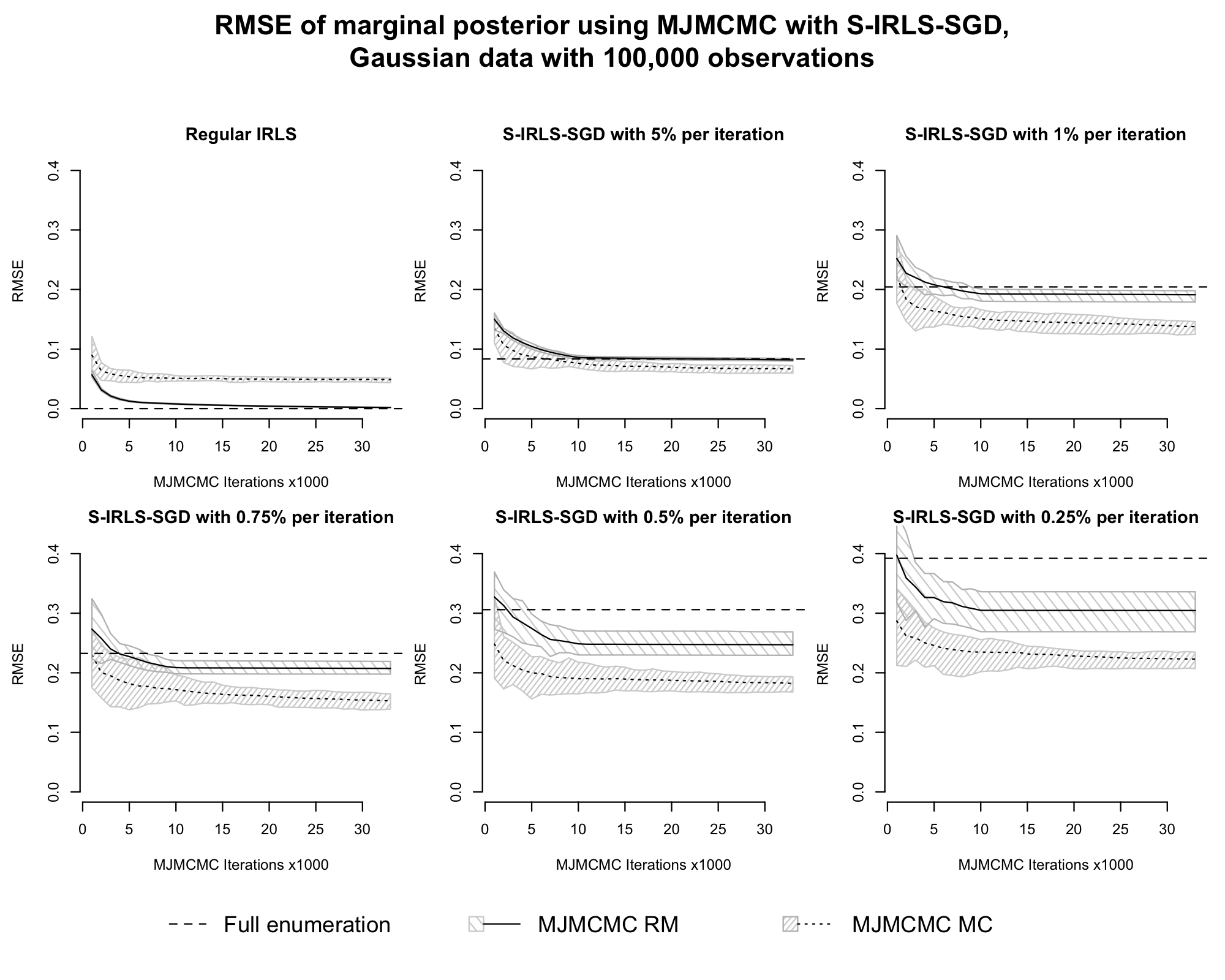}
	\caption{Root mean squared error compared to the true distribution for MJMCMC using both regular IRLS and S-IRLS-SGD with varying subsample sizes. MJMCMC was run 20 times and for 33,000 iterations for every variant of the algorithm. The black solid and dotted lines show the mean RMSE for the 20 runs, with the shaded areas showing 90\% confidence intervals, for RM and MC estimates respectively. The dashed line shows the RMSE for the full enumeration using the same algorithm settings.}
	\label{fig:mjmcmc_100KG}
\end{figure}

To be able to assess how well the algorithm converges, the RMSE was calculated for the first $1,000$, $2,000$, ..., $33,000$ iterations. The results for the data with $100,000$ observations are presented in Figures~\ref{fig:mjmcmc_100KG} and \ref{fig:mjmcmc_100KL}. Additionally, Figures~\ref{fig:mjmcmc_10Kg} and \ref{fig:mjmcmc_10Kl} in Appendix~\ref{add_results} contain the results for the data with $10,000$ observations. In all of the figures, we also present the RMSE for the full enumeration calculated using the same settings for S-IRLS-SGD.
\begin{figure}[h]
	\centering
	\includegraphics[scale=0.2]{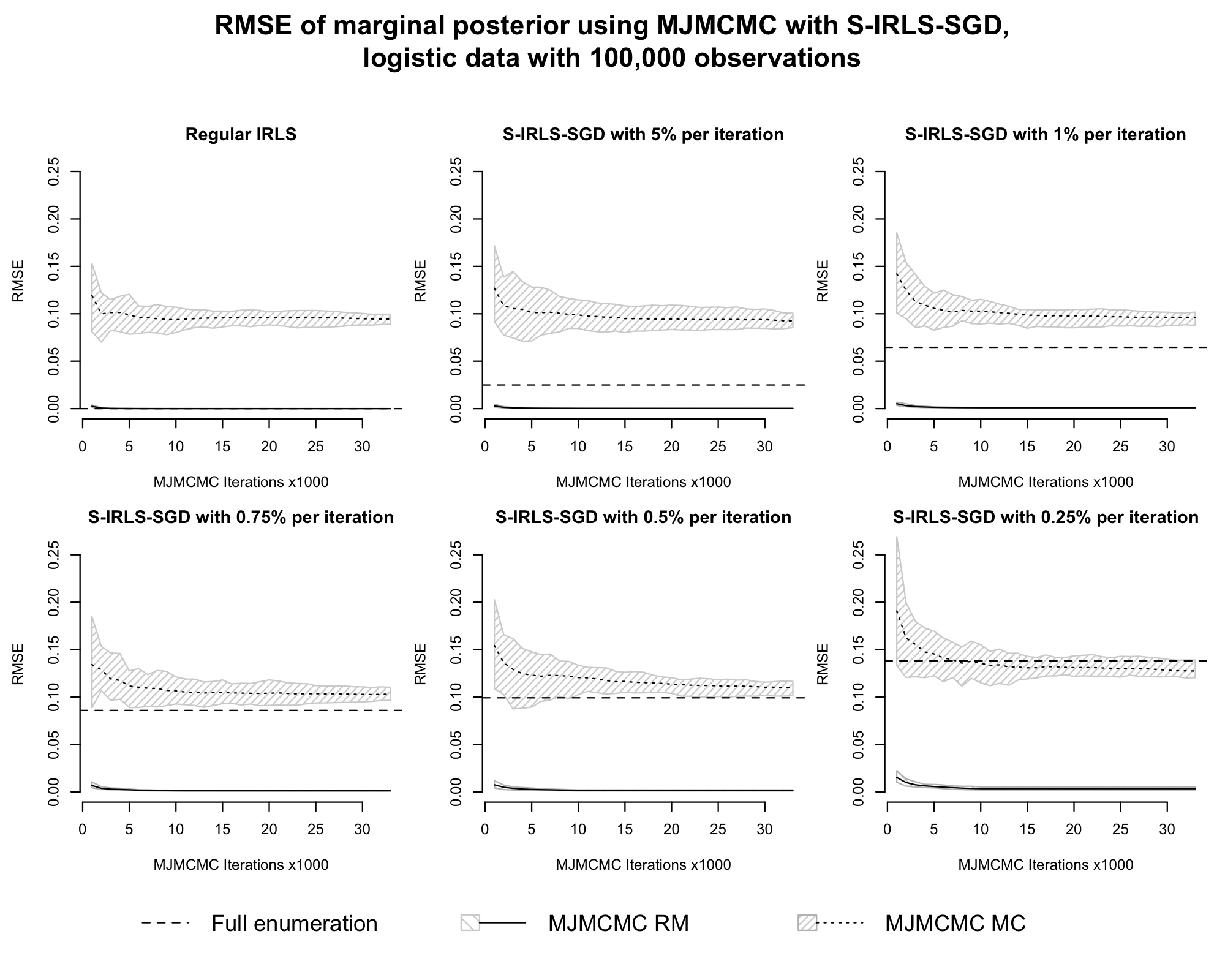}
	\caption{See caption of Figure~\ref{fig:mjmcmc_100KG}. Note that here the RM estimates have a very small error, for which reason they almost coincide with the $x$-axis in the plots.}
	\label{fig:mjmcmc_100KL}
\end{figure}

Regarding the results for the Gaussian data, we see in Figures~\ref{fig:mjmcmc_100KG} and \ref{fig:mjmcmc_10Kg} that the MC estimates perform better than their RM counterparts when using subsampling. Further, the full enumeration performs \textit{worse} than both estimates for all subsample sizes except for 5\%. There, the MC estimate outperforms full enumeration after around $5,000$-$10,000$ iterations while the RM estimate roughly coincides with it from around $10,000$. Another thing to note is that the convergence for the RM estimates appears to become very slow after an initially good performance. The same thing is noticed in the MC estimates, although not as pronounced.  Remarkably, the acceptance probability does not appear to drop significantly as the number of iterations increases.

The results obtained for the logistic models show a different pattern. In Figures~\ref{fig:mjmcmc_100KL} and \ref{fig:mjmcmc_10Kl}, we see that for every subsample size, the RM estimates perform very well, converging very close to the true distribution after very few iterations. On the other hand, the MC estimates quickly attain an RMSE of between 0.1-0.15 after which the convergence is very slow. Here, however, it should be noted that the approximations of the marginal likelihoods are used, i.e. Laplace approximations do not yield exact results under Bernoulli observations. Looking at the results as a whole, we can, as expected, see that a larger subsample provides a better estimate, albeit at a higher computational complexity per iteration.

\section{Discussion}\label{section5}

In this paper, we presented a novel approach to combine a BSGD algorithm for computing the marginal likelihood and MCMC/MJMCMC algorithms for Bayesian model selection. We also present a version of BSGD called S-IRLS-SGD. The S-IRLS-SGD algorithm has been implemented as a package in the programming language R with computationally intense blocks written in C++. It is publicly available at \url{https://github.com/jonlachmann/irls.sgd}.  To evaluate the performance of the S-IRLS-SGD algorithm to compute MLIK, we performed the experiments in Example~\hyperref[example1]{1}. There, we compared the performance of S-IRLS-SGD with that of IRLS for obtaining a posterior distribution of the components of $ \mathfrak{m} $ by enumerating every model in the model space. The results obtained show that we are able to get an acceptable level of precision given a rather small subsample size, encouraging us to continue on the chosen path. By selecting the highest estimate per model out of 5, 10 and 20 runs we observed that more evaluations result in even better estimates. With this, we were able to show empirical support to motivate Theorem~\ref{theorem2}, i.e. that multiple evaluations of a given model under S-IRLS-SGD provide convergence towards the true posterior distribution. The fact that we were both able to confirm that the algorithm worked and that the theory appeared to hold in practice encouraged us to continue. MJMCMC visits models of higher importance more often, which implies that the estimates for these models converge towards their true marginal likelihood faster than for any arbitrary model. Further, this means that under S-IRLS-SGD for computing MLEs, MJMCMC is able to provide \textit{even better} results than full enumeration using the same approximate optimisation algorithm.

Regular IRLS converge in two iterations for Gaussian models, and commonly in 7-20 iterations for logistic ones. For this reason, we chose to allow 20 S-IRLS iterations for initialisation and 250 BSGD iterations for final optimisation in the Gaussian experiments and $75+500$ iterations for logistic models. The S-IRLS iterations have a relative complexity of $ \mathcal{O}(n_s/n) $ compared to regular IRLS (for $ n \gg m $). The relative complexity of BSGD is $ \mathcal{O}(n_s/(nm)) $ compared to IRLS. For every subsample size used in the experiments, except 5\%, we achieve a lower complexity than regular IRLS. The proposed variant of BSGD (S-IRLS-SGD), is designed as an alternative to IRLS to ease the computational burden when estimating the MLE of a GLM. There is extensive literature and knowledge in the mathematical field of different optimisation algorithms, and we are sure that there are a lot of possible improvements to be made. More specifically, various extensions to SGD are available, possibly providing better global convergence properties for ill-conditioned likelihood functions as discussed by \citet{wedderburn1976existence}. For example, modern variants of SGD optimisation like Adam, Adagrad, RMSprop or others from \citet{bottou2018optimization} could be addressed instead of the basic BSGD algorithm or S-IRLS-SGD. Yet, the focus of the paper was not to find the best performing SGD method in combination with MCMC/MJMCMC, but rather to give principal novel ways to combine a general SGD and MCMC/MJMCMC for computationally efficient Bayesian model selection.  

\begin{figure}[h]
	\centering
	\includegraphics[scale=0.28]{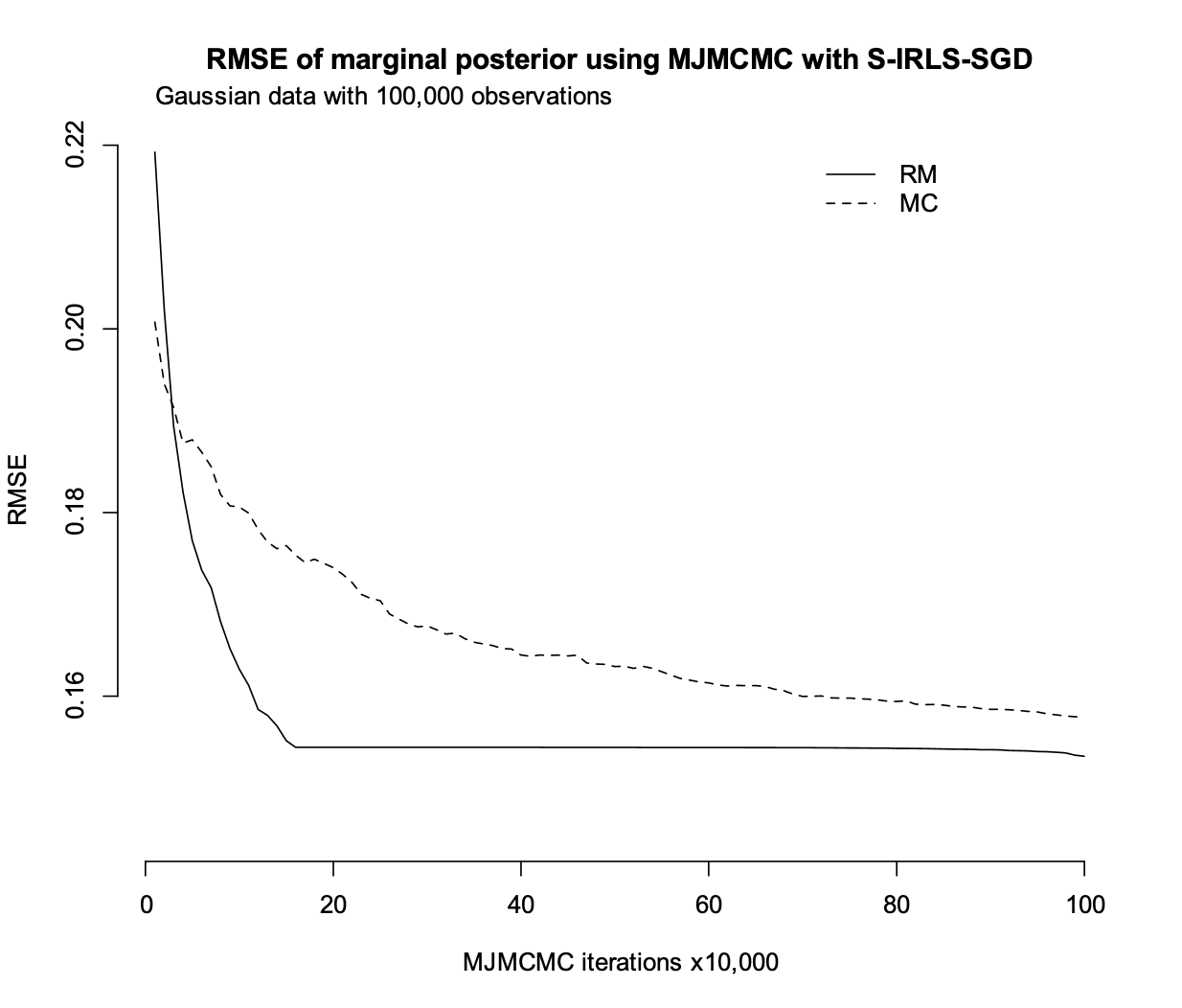}
	\caption{Root mean squared error compared to the true distribution for MJMCMC using S-IRLS-SGD with a subsample size of 0.75\%. MJMCMC was run for 1,000,000 iterations. The dashed lines shows the RMSE for the MCMC estimates and the solid lines shows RMSE for the renormalised ones.}
	\label{fig:discussion}
\end{figure}

In Section~\ref{sirlssgd_mjmcmc}, we presented three Theorems outlining possible strategies for combining BSGD (SGD, S-IRLS-SGD) and MCMC (MJMCMC). Throughout the examples, we relied on Theorem~\ref{theorem3}. By reestimating models at each visit and always selecting the best estimated marginal likelihood, it states that the estimate will converge to the exact Laplace approximation of the marginal likelihood. For Gaussian observations and certain types of priors, the approximation coincides with the exact marginal likelihood. In Example~\hyperref[example2]{2}, we combined S-IRLS-SGD with our implementation of MJMCMC through Theorem~\ref{theorem3}. The experiments were carried out using the same simulated data as in Example~\hyperref[example1]{1}. Here, we confirm that MJMCMC in most cases is able to provide a better estimate of the posterior marginal inclusion probabilities than full enumeration with the same settings for S-IRLS-SGD. For the logistic models, the RMSE of the RM estimates is very low, meaning that we are very close to the true distribution, even when we were just using 0.25\% of the data at each iteration. For Gaussian data, S-IRLS-SGD in MJMCMC perform significantly worse, probably owing to the smaller budget for iterations used. Yet, a convergence trend was still numerically confirmed.

The fact that we obtained a \textit{better} estimate using MJMCMC than full enumeration is a testament to its usefulness. It also provides empirical support for Theorem~\ref{theorem3}. In a setting where the number of models is not possible to enumerate in a reasonable amount of time, the only alternative is to use some sort of search algorithm. As \citet{hubin2018mjmcmc} have shown, MJMCMC is competitive in that field, performing on par or better than leading algorithms such as BAS by \citet{clyde2011bas}. In the case of logistic models, a higher iteration count was allowed for S-IRLS-SGD. This resulted in a higher computational complexity in absolute terms, but still compared well to the complexity of the more numerous iterations of IRLS. Here, our algorithm performed so well that it should be possible to evaluate even smaller subsample sizes than presented in this paper. This would allow for even higher performance gains when the number of observations is very large.

However, in some cases, the empirical rate of convergence of the implemented approach can deteriorate drastically after the algorithm is run for long enough. To illustrate this, we tried to run the combined algorithm corresponding to Theorem~\ref{theorem3} for a significantly larger amount of iterations than in the examples. The obtained results for one run with one million iterations (with a subsample size of 0.75\% of the data having $100,000$ observations) using the Gaussian data and model from Example~\hyperref[example2]{2} are presented in Figure~\ref{fig:discussion}. There, we clearly see that the convergence becomes quite slow after a while, yet even at around one million iterations the algorithm continues to converge to the truth. Hence, additional theoretical studies of the convergence rates are required in the future to understand why convergence often becomes slower when the algorithm reaches a certain number of iterations. At the same time, it would be interesting to investigate if this slow convergence significantly influences the power and false discovery rate when doing Bayesian variable selection using the median probability model or the most probable model. We however, keep this as a potential subject for another article. Also, studying if the strategy that restarts the optimisation algorithm from the previous estimates when revisiting a model (from Theorem~\ref{theorem2}) improves the convergence properties of the combined algorithm is of interest for further research.

Lastly, as an alternative to the Laplace approximation for models with latent Gaussian structures, \textit{Integrated Nested Laplace Approximations (INLA)} have emerged as an efficient approximation method \citep{rue2009inla}. \citet{hubin2016estimating} showed empirically that INLA performs better than Laplace approximation in various scenarios. It would, thus, be interesting to extend the subsampling ideas put forth in this article to that context by combining them with INLA. Another interesting idea for future research is to apply the proposed methods of \citet{quiroz2018subsampling} when doing the final calculation of the deviance. This would further improve the speed of the only step in the current algorithm that depends on the full set of data. Yet, the methods from  \citet{quiroz2018subsampling} would break the regularity conditions of our Theorems~\ref{theorem2} and ~\ref{theorem3}, which means that additional theoretical studies would also be required to be able to incorporate it into our framework.

\bibliographystyle{chicago}
\bibliography{references}  
\newpage
\appendix

\section{Limitations and additional methodological contributions}\label{ap:A0}

\subsection{Limitations}

The likelihood function of any GLM is often believed to be concave with respect to the parameters, suggesting that any first or second-order optimiser should be able to find the MLE. This is however not always the case. \citet{wedderburn1976existence} define some properties for the MLE of various GLMs. Most importantly, they note that for some cases the MLE does not exist, is not unique, or is not finite. All of the versions of SGD and BSGD utilise gradient descent to ensure convergence and are therefore limited to having a theoretical guarantee only for solving concave problems.

We will here mention some important examples where these conditions are met, see \citet{wedderburn1976existence} for a complete list. Gaussian regression with identity link is guaranteed to have a finite and unique MLE that exists in the interior of the parameter space. Logistic regression with \textit{logit} or \textit{probit} link has a well-behaved MLE in terms of uniqueness and existence in the interior of the parameter space. It is however only finite as long as degenerate cases of perfect separation are excluded.

As our testing was limited to Gaussian and logistic models, we will only present empirical proof of the convergence for these. It should also be noted that the simulated data in the logistic examples were generated in a way  to avoid perfect separation. From \citet{robbins1971proof} and \citet{bottou2018optimization}, we have the result that SGD, BSGD, and consecutively S-IRLS-SGD will eventually converge to the optimum of a concave function. Hence,  at least theoretically, BSGD will converge in every case where the MLE exists, is unique and finite.

\begin{figure}[h]
    \centering
    \includegraphics[scale = 0.2]{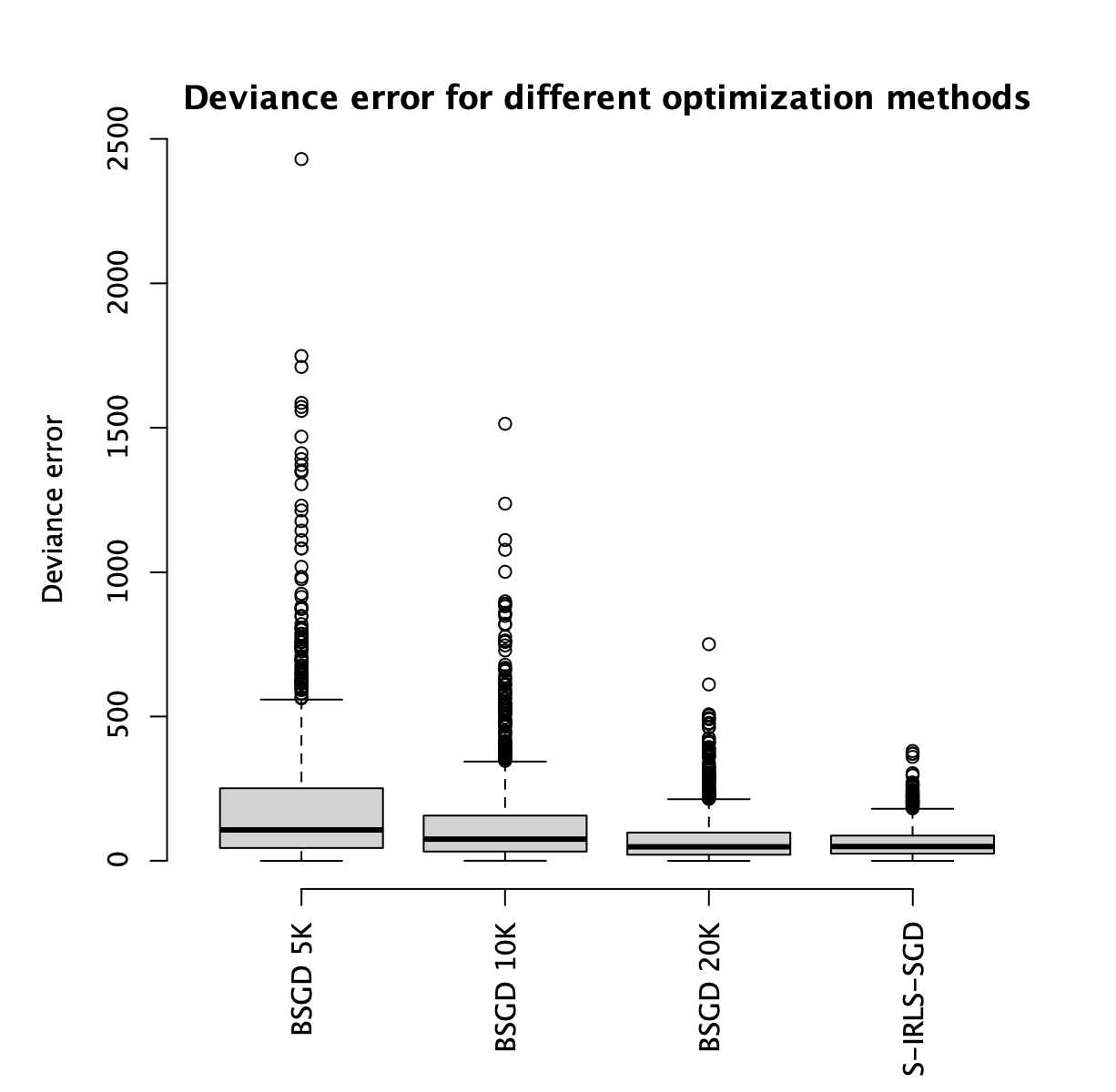}
    \caption{Boxplots of the errors of estimated deviances for the  last four optimisation algorithms and tuning parameters from Table \ref{table:optim} and Figure~\ref{fig:sgdserr}.} \label{fig:sgdserr2}
\end{figure}

\subsection{Subsampling IRLS}\label{sec:S-IRLS}
For IRLS, the computational complexity is dominated by the matrix multiplication and matrix inversion in the weighted least squares (WLS) step which has complexity $ \mathcal{O}(\max(m^2n, m^3)) $, where $n$ is the number of observations and $m$ is the number of parameters to be estimated. Hence, for a set of data with many observations, this complexity will be dominated by $ \mathcal{O}(m^2n) $ which grows linearly with the number of observations. To mitigate this, we have developed a variant of IRLS called \textit{Subsampling IRLS (S-IRLS)} which uses only a small portion of the data in each iteration, reducing the complexity significantly.

In each iteration a random subsample is drawn with selection probabilites proportional to $ \bm w +  \bm\varepsilon_{w}$, where $ \bm w $ is the current weight vector and $ \bm\varepsilon_{w} $ is used to regularise the sampling procedure and make sure that each observation can be sampled. This way, we can estimate the expected information matrix $ \bm x^T \text{diag}(\bm w) \bm x $ on a subsample and the complexity is reduced to  $ \mathcal{O}(\max(m^2n_s, m^3)) $, where $ n_s $ is the size of our subsample. The algorithm is described formally as Algorithm~\ref{sub_irls}.

\begin{algorithm}[H]
	\SetAlgoLined
	Let $ t = 1 $.\\
	Initialise the weights vector $\bm w$ and the mean vector $\bm \mu_{\text{start}}$ for the current GLM family.\\
	Select a random subsample $ \mathcal{S} \subset  \left\lbrace 1,...,n \right\rbrace $, such that $||\mathcal{S}|| = n_\mathcal{S}$.\\
	Calculate the linear predictor and mean vector
	$ \bm \eta = \mathsf{h}(\bm \mu_{\text{start},\mathcal{S}}) $, $ \bm \mu = \mathsf{h}^{-1}(\bm \eta) $.\\
	\While{$ t \leq T $}{
		Let:\\
		The variance of the mean vector $ \sigma^2_{\bm \mu} = \text{Var}(\bm \mu) $,\\
		The derivative of the inverse link w.r.t. the linear predictor $ \bm \xi = \frac{d \mathsf{h}^{-1} (\bm \eta)}{d \bm \eta} $,\\
		The working response vector $ \bm z = \bm \eta + (\mathbf{y}_\mathcal{S} - \bm \mu) / \bm \xi, $\\
		Update the weights for the current subsample $ \bm w_{\mathcal{S}} = \left( {\bm \xi^2}/{\sigma^2_{\bm \mu}} \right)^{0.5} $.\\
		Solve the Weighted Least Squares equation
		\begin{align}
			\bm \beta_t^{\star} = (\bm x_{\mathcal{S}}^T \text{diag}(\bm w_\mathcal{S}) \bm x_\mathcal{S})^{-1} \bm x_\mathcal{S}^T \text{diag}(\bm w_\mathcal{S}) \bm z.
		\end{align}\\
		Update the values of $ \bm \beta_t $ according to the cooling schedule
		\begin{align}
			\bm \beta_t = \tau_t \bm \beta_t^{\star} + (1- \tau_t)\bm \beta_{t-1}. \label{eq:cooling}
		\end{align}\\
		Update the subsample by selecting elements $i \in \mathcal{S}$ with probabilities proportional to  $w_i + \bm\varepsilon_{w}$, where $\bm w$ is the vector of weights from the previous step of the algorithm.\\
		Calculate the new linear predictor and mean vector $ \bm \eta = \bm x_\mathcal{S} \bm \beta_t $, $ \bm \mu = \mathsf{h}^{-1}(\bm \eta) $.\\
		Calculate the estimated deviance $ d_t $ based on the current subsample $\mathcal{S}$.\\
		\If{$ (d_t - d_{t-1}) / |d_{t-1}| > \delta_{\text{expl}} $}{
			Reset the values of $ \bm \beta_t $ to those from two iterations ago and halve the temperature:
			\begin{align}
				\bm \beta_t = \bm \beta_{t-2}, \quad \tau_t = \frac{\tau_t}{2}.
			\end{align}\\
			Recalculate the new linear predictor and mean vector $ \bm \eta = \bm x_\mathcal{S}^T \bm \beta_t $, $ \bm \mu = \mathsf{h}^{-1}(\bm \eta) $.\\
		}
		Let $ t = t+1 $.
	}
	\caption{Subsampling Iteratively Reweighted Least Squares (S-IRLS).}
	\label{sub_irls}
\end{algorithm}

S-IRLS requires the user to set a number of tuning parameters for optimal performance, these are
\begin{itemize}
	\item $n_s $ \dots size of subsamples
	\item $ T $ \dots total number of iterations 
	\item $ \tau_0, \tau_d, t_{\text{const}} $ \dots three parameters which determine the temperature $\tau_t$ in equation~\eqref{eq:cooling}
	\item $\delta_{\text{expl}} $ \dots threshold on the relative increase of deviance 
\end{itemize}

The temperature $ \tau $ which is used to improve the convergence properties of the algorithm determines the rate at which $ \bm \beta $ is allowed to change between the iterations, see Equation \eqref{eq:cooling}. We have chosen to use what is called a constant - exponential decay cooling schedule, which means that the temperature is kept constant for some initial iterations determined by $t_{\text{const}}$, after which it is subject to exponential decay
\begin{align}
	\tau_t = \tau_0 \cdot \tau_d^{\max((t - t_{\text{const}}),0)}.\label{eq:cooling2}
\end{align}
This kind of cooling is also found in some SGD implementations to homogenise the iterative solutions and avoid an unreasonably large impact from an unfortunate subsample. S-IRLS is able to quickly approach the vicinity of the MLE $\bm{\hat \beta}$ even when the starting point is far from it.  However, when approaching the MLE, the estimated Fisher information tends to become very unstable and there is the risk that the algorithm might not converge at all. This problem is taken into account by detecting what we call \textit{exploding deviance}, which we define to be the case when the estimated deviance increases by a factor larger than  $\delta_{\text{expl}}$. This happens more often  the smaller the size of the subsample, because then the estimate of the expected Fisher information will be more unstable.  If this occurs, following popular IRLS implementations for GLM, the current estimate $\bm \beta$ is set to $\bm \beta_{t-2}$ and the temperature is halved to avoid a loop of deviance explosions and backtracking. 
 
Finally, it is important to note that S-IRLS  is (at least for the time being) a heuristic procedure and, in general, does not have any guarantee to converge to an optimal solution if applied on its own. Hence, we only use S-IRLS to estimate the starting solutions for BSGD, which is run to ensure convergence.

\section{Reproducing previous results for MJMCMC}\label{A1}
To verify that the implementation of MJMCMC contributed in this paper works as intended, some previous results from \citet{hubin2018mjmcmc} were reproduced and compared against the original implementation. Examples \hyperref[example1]{1} and \hyperref[example2]{2} compare the performance of the new implementation of MJMCMC with the previous one using real and simulated data, the first example corresponds to Example 1 in \citet{hubin2018mjmcmc} and the second one corresponds to Example 2 of the same paper.

\subsubsection*{Example A1: U.S. Crime Data} \label{exampleA1}
The \textit{U.S. Crime Data Set} compiled by \citet{Vandaele1992} contains crime rates for 47 U.S. states along with 15 covariates. Following \citet{hubin2018mjmcmc}, we will test our implementation of MJMCMC against it to verify that the algorithm performs as it should. With one dependent variable and 15 independent, it is possible to create $ 2^{15} $ different models, including one with only an intercept. \citet{hubin2018mjmcmc} followed \citet{clyde2011bas} in using Zellner's g-prior \citep{zellner1986bayesian} with $ g=47 $, something we will also do. This gives us the exact marginal likelihood
\begin{align}
    p(\mathbf{y} | \mathfrak{m}) \propto (1+g)^{(n-p-1)/2}(1+g[1-R^2_{\mathfrak{m}}])^{-(n-1)/2},
\end{align}
where $ R^2_{\mathfrak{m}} $ is the coefficient of determination for the model $ \mathfrak{m} $.

Since the total number of possible models is reasonably small, we began by enumerating all of them to calculate the true marginal distribution of the components of the $ \mathfrak{m} $ vector. We then ran MJMCMC on the data 20 times for a number of iterations to visit approximately the same number of unique models as in the results we wanted to replicate. The reason for the approximation is that the previous implementation of MJMCMC is configured such that it runs until it encounters a fixed number of either unique or total models, whereas our implementation runs for a fixed number of iterations. For every parameter estimate, we calculated the RMSE for each indicator variable $\gamma_j$ as defined in Equation~\eqref{eq:rmse}.

The true values and the RMSE multiplied by 100 of the renormalized (RM) and MCMC (MC) estimates are presented in Table~\ref{table:crime1}, along with the corresponding values from \citet{hubin2018mjmcmc}. The posterior mass captured by MJMCMC is also shown there.
We can see that our results for the RM estimates are very similar to those previously reported, and the MC estimates are a bit better than the ones obtained in the original implementation. It should however be noted that the total number of models visited in our runs is a bit higher, something caused by the different ways the two implementations of the algorithm are tuned. For a comparison to competing algorithms such as BAS, MC$^3$ and RS, refer to Table 3 of \citet{hubin2018mjmcmc} and to \citet{clyde2011bas}.

\begin{figure}
	\centering
	\includegraphics[scale=0.2]{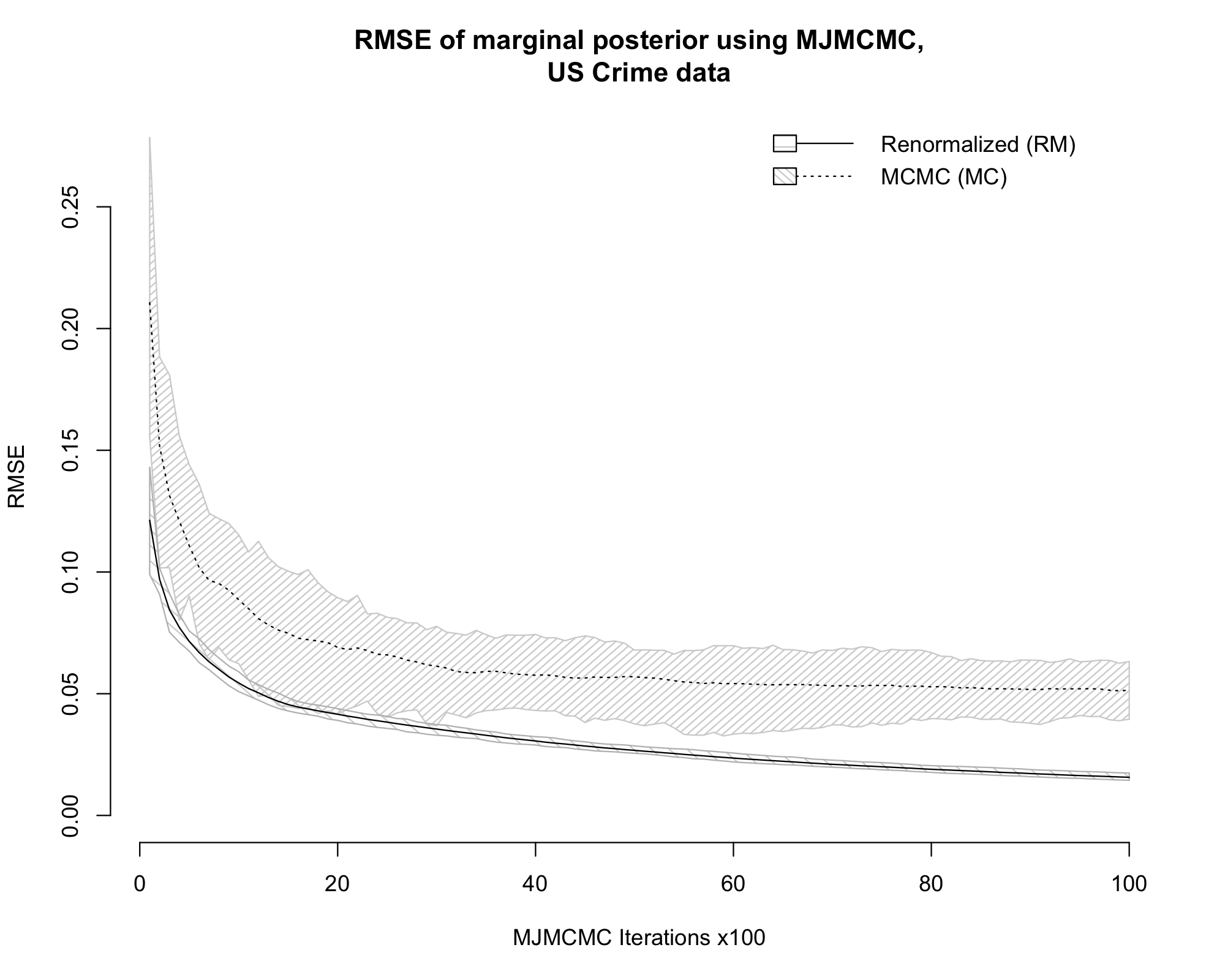}
	\caption{RMSE compared to the true distribution for MJMCMC run with the US Crime data in Example~\hyperref[exampleA1]{A1}. MJMCMC was run 20 times and for $10,000$ iterations. The black solid and dotted lines show the mean RMSE for the 20 runs, with the shaded areas showing 90\% confidence intervals, for RM and MC estimates respectively.}
	\label{fig:crime}
\end{figure}

To demonstrate the convergence of the algorithm, we also show the RMSE plotted against the number of iterations for both RM and MC estimates in Figure~\ref{fig:crime}. We note here that the RM estimates converge quicker and more stably towards the true distribution for all of the components.

\begin{table} \centering
	\caption{$ \text{RMSE} \times 100 $ for RM and MC estimates of $ \mathfrak{m} $ marginal inclusion probabilities based on $ 2 \times 20 $ different runs of MJMCMC in Example~\hyperref[exampleA1]{A1}. Also shown is the mean RMSE and total posterior mass captured. Previous results reported in \citet{hubin2018mjmcmc} in parentheses.} 
	\label{table:crime1}
	\vspace{5mm}
	\begin{tabular}{c|c|cc|cc}
  & True value & RM & MC & RM & MC \\
 \hline
$\gamma_8$      & 0.16 & 7.76 (6.57) & 6.27 (10.68) & 6.22 (5.11) & 6.19 (10.29) \\
$\gamma_{13}$   & 0.16 & 7.42 (7.46) & 6.32 (10.54) & 6.04 (5.60) & 5.96 (10.19) \\
$\gamma_{14}$   & 0.19 & 8.62 (8.30) & 6.33 (12.43) & 7.01 (6.30) & 6.26 (12.33) \\
$\gamma_{12}$   & 0.22 & 7.96 (6.87) & 8.71 (13.61) & 6.56 (5.57) & 6.25 (13.64) \\
$\gamma_5$      & 0.23 & 7.69 (6.30) & 6.51 (13.45) & 5.76 (4.59) & 5.64 (13.65) \\
$\gamma_9$      & 0.23 & 9.28 (9.49) & 7.18 (16.21) & 7.08 (7.40) & 6.87 (16.21) \\
$\gamma_7$      & 0.29 & 7.10 (4.37) & 4.84 (13.63) & 4.93 (3.45) & 3.36 (12.73) \\
$\gamma_4$      & 0.30 & 5.66 (6.18) & 8.75 (19.22) & 3.58 (3.79) & 6.27 (17.31) \\
$\gamma_6$      & 0.33 & 8.26 (8.61) & 6.94 (19.71) & 5.53 (6.14) & 4.55 (19.49) \\
$\gamma_1$      & 0.34 & 9.40 (11.32) & 7.05 (22.68) & 7.10 (7.29) & 6.17 (20.50) \\
$\gamma_3$      & 0.39 & 4.40 (3.95) & 9.93 (11.13) & 2.00 (2.38) & 7.41 (6.99) \\
$\gamma_2$      & 0.57 & 3.77 (5.92) & 13.97 (13.21) & 2.93 (3.82) & 11.00 (9.03) \\
$\gamma_{11}$   & 0.59 & 3.81 (3.57) & 9.61 (13.49) & 1.43 (2.37) & 5.96 (15.94) \\
$\gamma_{10}$   & 0.77 & 6.44 (7.62) & 13.32 (7.28) & 5.23 (5.97) & 10.94 (4.78) \\
$\gamma_{15}$   & 0.82 & 5.98 (9.23) & 14.51 (4.45) & 5.04 (6.89) & 12.34 (5.85) \\
\hline
 & Mean RMSE & 6.90 (7.05) & 8.68 (13.45) & 5.10 (5.11) & 7.01 (12.60) \\
 \hline
 & Total mass & 0.44 (0.58) &  & 0.58 (0.71) &  \\
 \hline
 & Unique models & ${1900}$ (1908) &  & $ {3219} $ (3237) &  \\
 \hline
 & Total models & 6200 (3276) &  & 11200 (5936) & 
\end{tabular}
\end{table}

\subsubsection*{Example A2: Simulated Data for Logistic Regression (MJMCMC)} \label{exampleA2}
\begin{figure}
	\centering
	\includegraphics[scale=0.175]{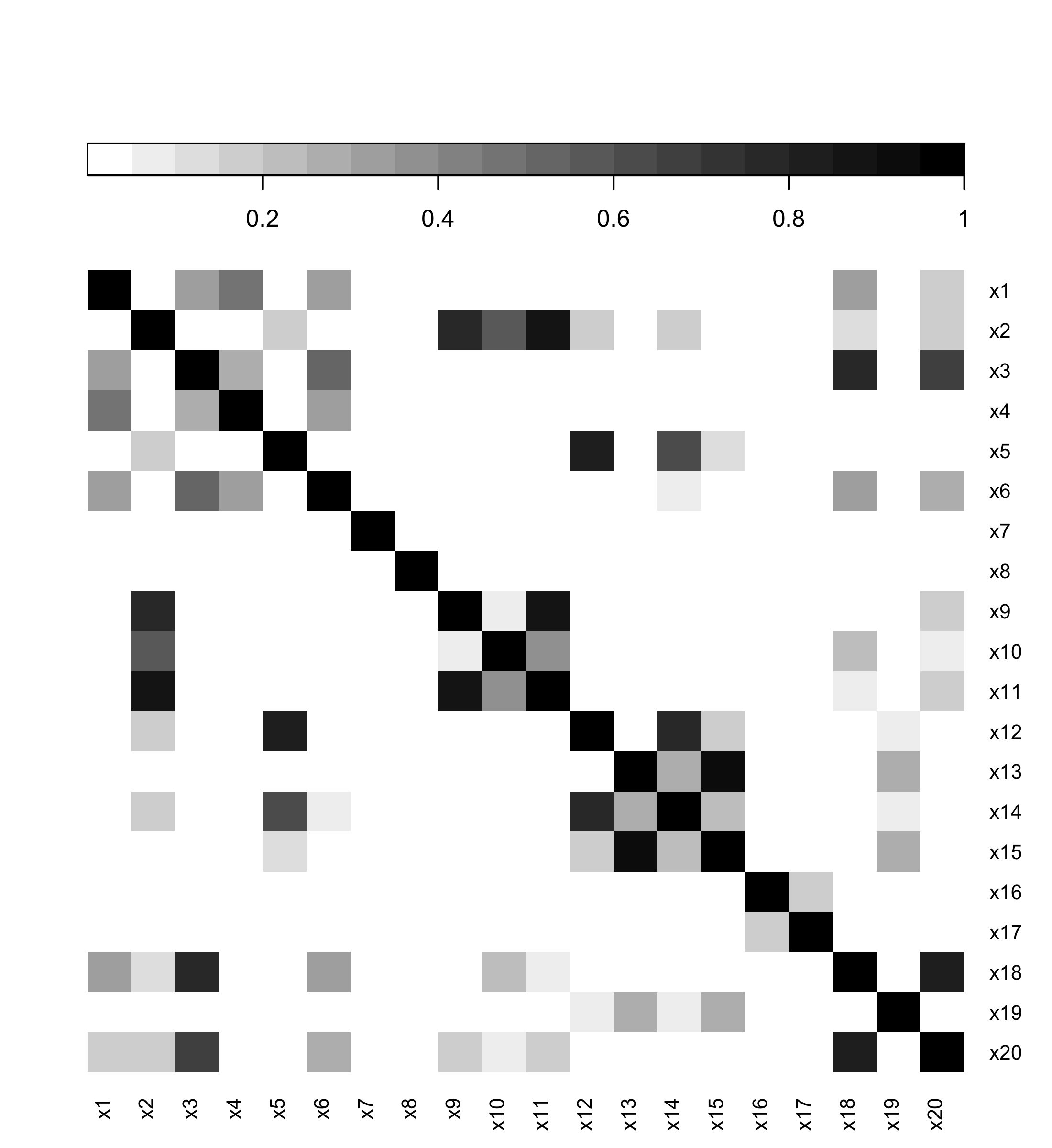}
	\caption{Absolute correlation between the exogenous $ x $ variables for the data used in Example~\hyperref[example2]{2}. This complicated correlation structure ensures that the distribution of models is multi-modal.}
	\label{fig:ex2_cor}
\end{figure}

\citet{hubin2018mjmcmc} created a dataset to be used for logistic regression where the model space is highly multi-modal, aiming to properly assess the convergence in this more complicated setting. The covariance structure is visualised in Figure~\ref{fig:ex2_cor}. The data consists of 20 independent variables $ \bm x $ and one dependent variable $ \mathbf{y} $, each with 2,000 observations. This results in $2^{20}=1,048,576$ possible models, providing a model space that is much larger and more computationally intensive to fully enumerate than in the previous example. This highlights the need for a good search algorithm to properly explore it without having to calculate every possible model. In this example, the number of observations is also larger. Hence, even the time it takes to calculate a single model becomes much higher than in the previous example.

\begin{figure}
	\centering
	\includegraphics[scale=0.2]{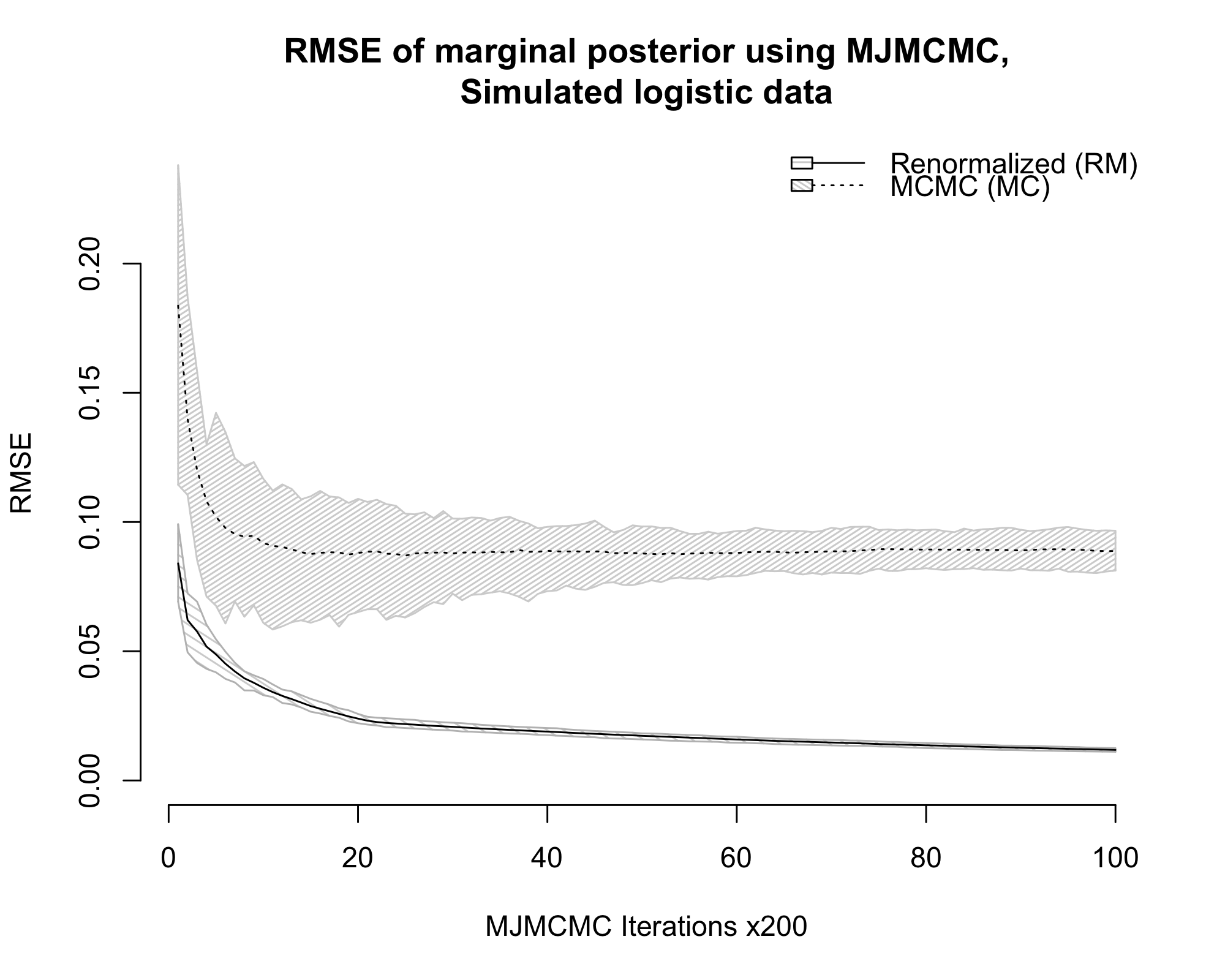}
	\caption{RMSE compared to the true distribution for MJMCMC run with the simulated logistic data in Example~\hyperref[exampleA2]{A2}. MJMCMC was run 20 times and here the first $20,000$ iterations are used to demonstrate the convergence. The black solid and dotted lines show the mean RMSE for the 20 runs, with the shaded areas showing 90\% confidence intervals, for RM and MC estimates respectively.}
	\label{fig:ex2_rmse}
\end{figure}

We used the AIC prior from \citet{clyde2020bas}  to be able to compare the results with those of \citet{hubin2018mjmcmc}. Using AIC, the marginal likelihood is approximated via the deviance $D(\mathbf{y})$ as
\begin{align}
    \hat p(\mathbf{y} | \mathfrak{m}) \propto -\frac{1}{2} \left( D(\mathbf{y}) + 2 \sum \gamma_i \right).
\end{align}
To provide a baseline to compare against we first performed a full enumeration of every possible model. This both gave us exact values and verified that our marginal likelihood was correctly implemented. We then ran MJMCMC 20 times for $32,768$ iterations to be able to extract subsets of these runs that are comparable to the previous results. The RMSE after visiting a similar number of unique models are presented in Table~\ref{table:ex2}, along with the true distribution and a comparison to previous results. In Figure \ref{fig:ex2_rmse}, we also demonstrate the convergence of the RM and MC estimates over the course of $20,000$ iterations.

By examining the results we see that the RM estimates are very similar to those of \citet{hubin2018mjmcmc}. Just as in the previous example, the MC estimates are slightly more accurate in our implementation. Also here it should be noted that while we estimated roughly the same number of unique models, we visited more models in total. Yet, we captured a bit less of the posterior mass than in the original implementation of MJMCMC. 

\begin{table} \centering
	\caption{Example~\hyperref[exampleA2]{A2}. $ \text{RMSE} \times 100 $ for RM and MC estimates of $ \mathfrak{m} $ parameters based on $ 2 \times 20 $ different runs of MJMCMC. Also shown is the mean RMSE and total posterior mass captured. Previous results reported in \citet{hubin2018mjmcmc} in parentheses.} 
	\label{table:ex2}
	\vspace{5mm}
\begin{tabular}{c|c|c|c|c|c}
 & True value & RM & MC & RM & MC \\
 \hline
$\gamma_6$      & 0.29 & 7.96 (7.38) & 7.20 (15.54) & 5.56 (4.54) & 7.20 (16.62) \\
$\gamma_8$      & 0.31 & 7.83 (6.23) & 5.47 (15.50) & 5.17 (3.96) & 5.48 (16.94) \\
$\gamma_{12}$   & 0.35 & 4.92 (4.86) & 8.99 (14.62) & 3.19 (2.78) & 8.37 (13.66) \\
$\gamma_{15}$   & 0.35 & 4.82 (4.55) & 10.92 (15.24) & 2.49 (2.56) & 11.73 (15.45) \\
$\gamma_2$      & 0.36 & 5.15 (4.90) & 6.88 (16.52) & 3.45 (2.92) & 7.08 (17.39) \\
$\gamma_{20}$   & 0.37 & 6.30 (4.82) & 6.90 (14.35) & 4.13 (2.66) & 6.90 (14.08) \\
$\gamma_3$      & 0.40 & 6.89 (9.25) & 12.20 (20.93) & 4.23 (5.65) & 12.87 (22.18) \\
$\gamma_{14}$   & 0.44 & 2.40 (3.14) & 3.76 (17.54) & 1.36 (1.58) & 2.84 (16.24) \\
$\gamma_{10}$   & 0.44 & 2.82 (4.60) & 6.35 (18.73) & 1.09 (2.29) & 6.13 (17.90) \\
$\gamma_5$      & 0.46 & 2.17 (3.10) & 4.07 (17.17) & 0.98 (1.53) & 2.73 (16.97) \\
$\gamma_9$      & 0.61 & 4.02 (3.68) & 3.76 (16.29) & 2.51 (1.63) & 3.07 (13.66) \\
$\gamma_4$      & 0.88 & 6.65 (5.66) & 16.30 (6.70) & 5.48 (3.74) & 16.13 (6.26) \\
$\gamma_{11}$   & 0.91 & 3.97 (5.46) & 14.27 (6.81) & 2.92 (3.95) & 14.13 (6.90) \\
$\gamma_1$      & 0.97 & 2.05 (1.90) & 13.46 (1.74) & 1.70 (1.35) & 13.21 (1.34) \\
$\gamma_{13}$   & 1.00 & 0.00 (0.00) & 11.31 (0.43) & 0.00 (0.00) & 11.13 (0.32) \\
$\gamma_7$      & 1.00 & 0.00 (0.00) & 6.88 (0.57) & 0.00 (0.00) & 6.79 (0.41) \\
$\gamma_{16}$   & 1.00 & 0.00 (0.00) & 6.37 (0.41) & 0.00 (0.00) & 5.99 (0.33) \\
$\gamma_{17}$   & 1.00 & 0.00 (0.00) & 6.25 (0.43) & 0.00 (0.00) & 6.11 (0.39) \\
$\gamma_{18}$   & 1.00 & 0.00 (0.00) & 6.31 (0.47) & 0.00 (0.00) & 6.19 (0.35) \\
$\gamma_{19}$   & 1.00 & 0.00 (0.00) & 6.82 (0.52) & 0.00 (0.00) & 6.78 (0.36) \\
\hline
 & Mean RMSE & 3.40 (3.48) & 8.22 (10.03) & 2.21 (2.06) & 8.04 (9.89) \\
 \hline
 & Total mass & 0.54 (0.72) &  & 0.73 (0.85) &  \\
 \hline
 & Unique models & 5151 (5148) &  & 9993 (9988) &  \\
 \hline
 & Total models & 13250 (9998) &  & 27200 (19849) & 
\end{tabular}
\end{table}

\newpage
\FloatBarrier

\section{Additional results for Example~\hyperref[example1]{1}}\label{add_results}
\begin{figure}[!ht]
	\centering
	\includegraphics[scale=0.16]{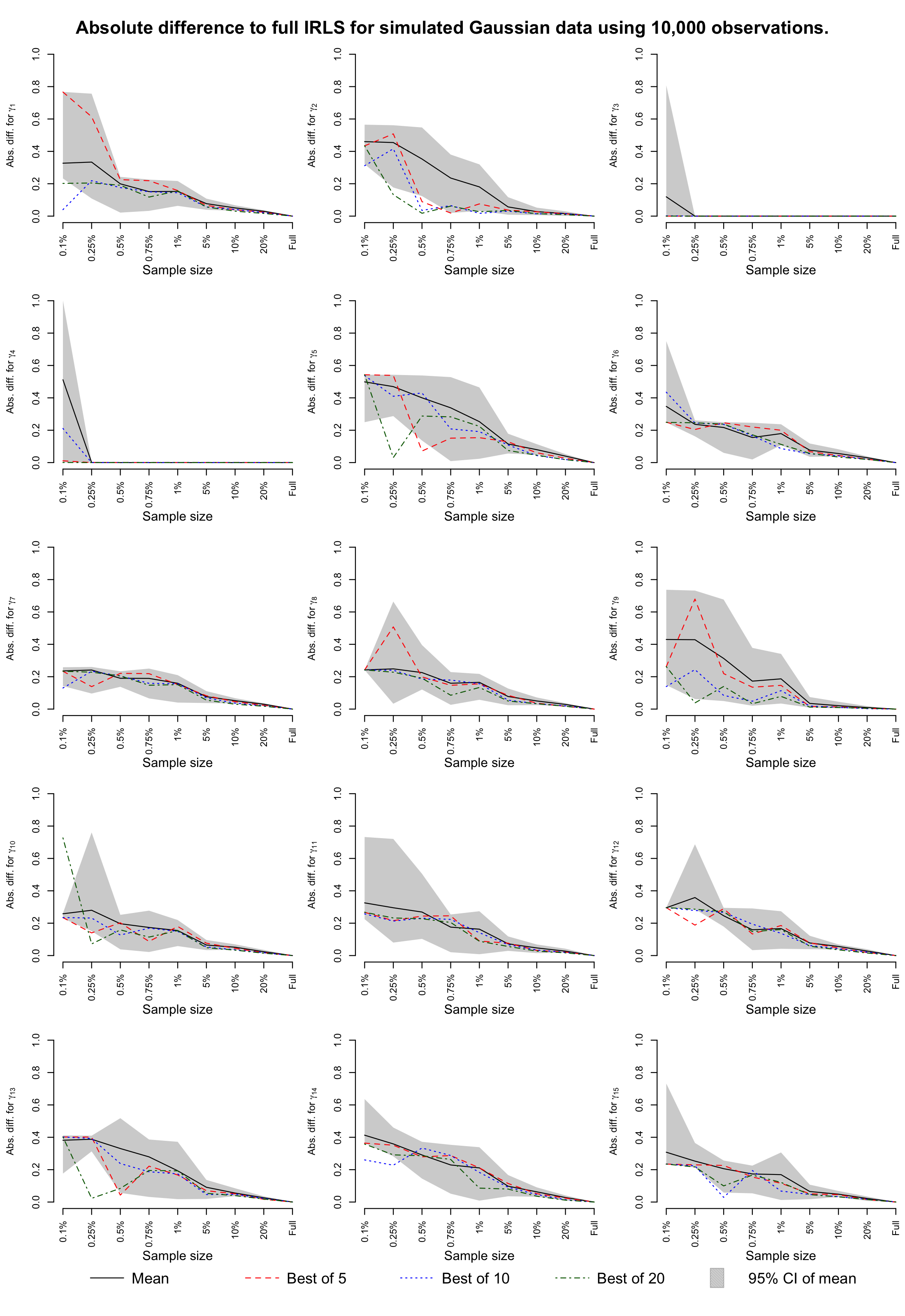}
	\caption{The absolute difference for marginal posterior probabilities for full enumeration of 32,768 Gaussian models using $10,000$ observations from the set of data in Example~\hyperref[example1]{1}. 20 runs were performed and the black line shows the mean absolute difference, with grey 90\% confidence intervals. The dashed and dotted lines show the absolute difference obtained by using the best likelihood estimate for each model that occurred during 5, 10 and 20 random runs.}
	\label{fig:full_10Kg}
\end{figure}
\begin{figure}
	\centering
	\includegraphics[scale=0.16]{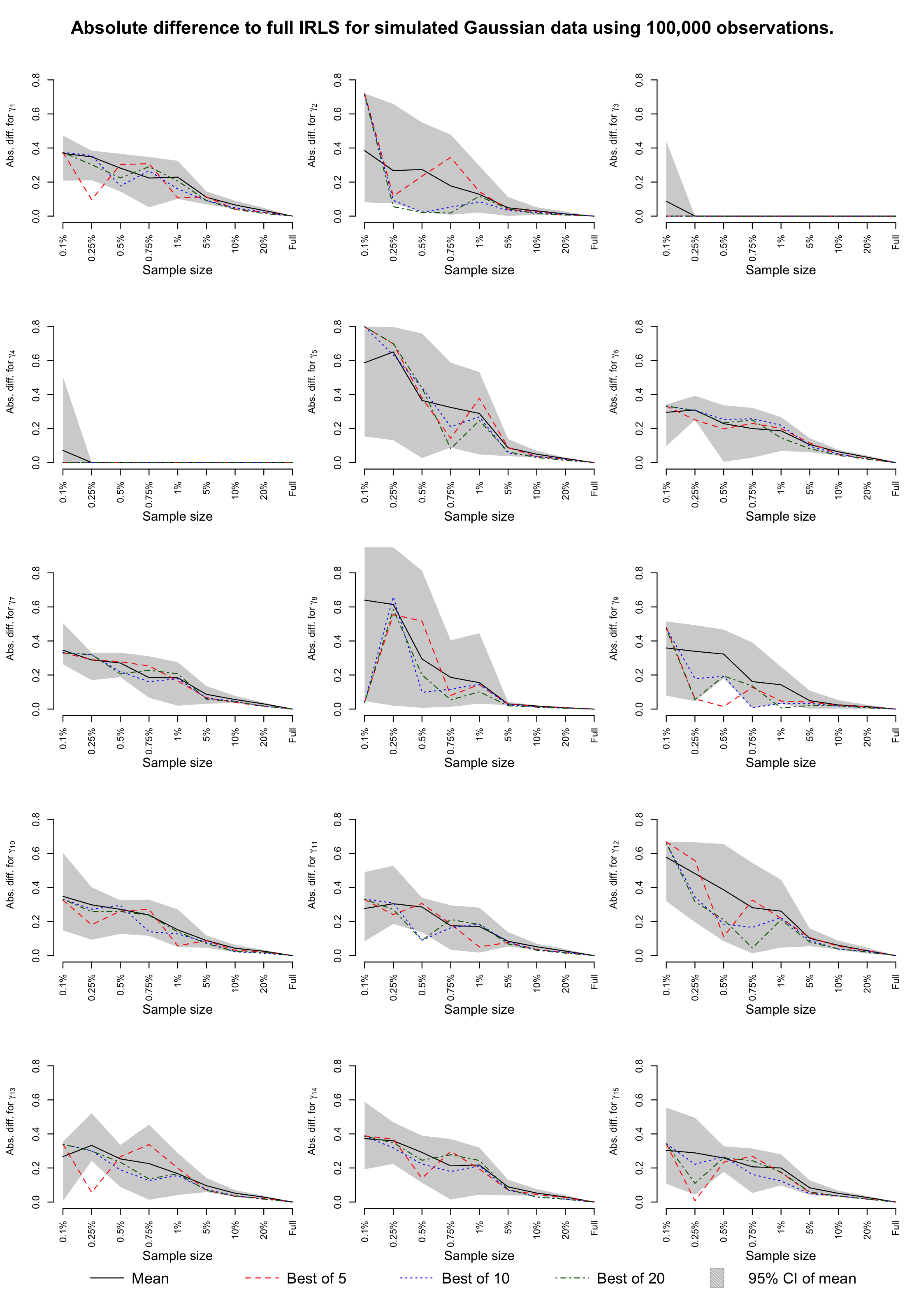}
	\caption{The absolute difference for marginal posterior probabilities for full enumeration of 32,768 Gaussian models using $100,000$ observations from the set of data in Example~\hyperref[example1]{1}. 20 runs were performed and the black line shows the mean absolute difference, with grey 90\% confidence intervals. The dashed and dotted lines show the absolute difference obtained by using the best likelihood estimate for each model that occurred during 5, 10 and 20 random runs.}
	\label{fig:full_100Kg}
\end{figure}
\begin{figure}
	\centering
	\includegraphics[scale=0.16]{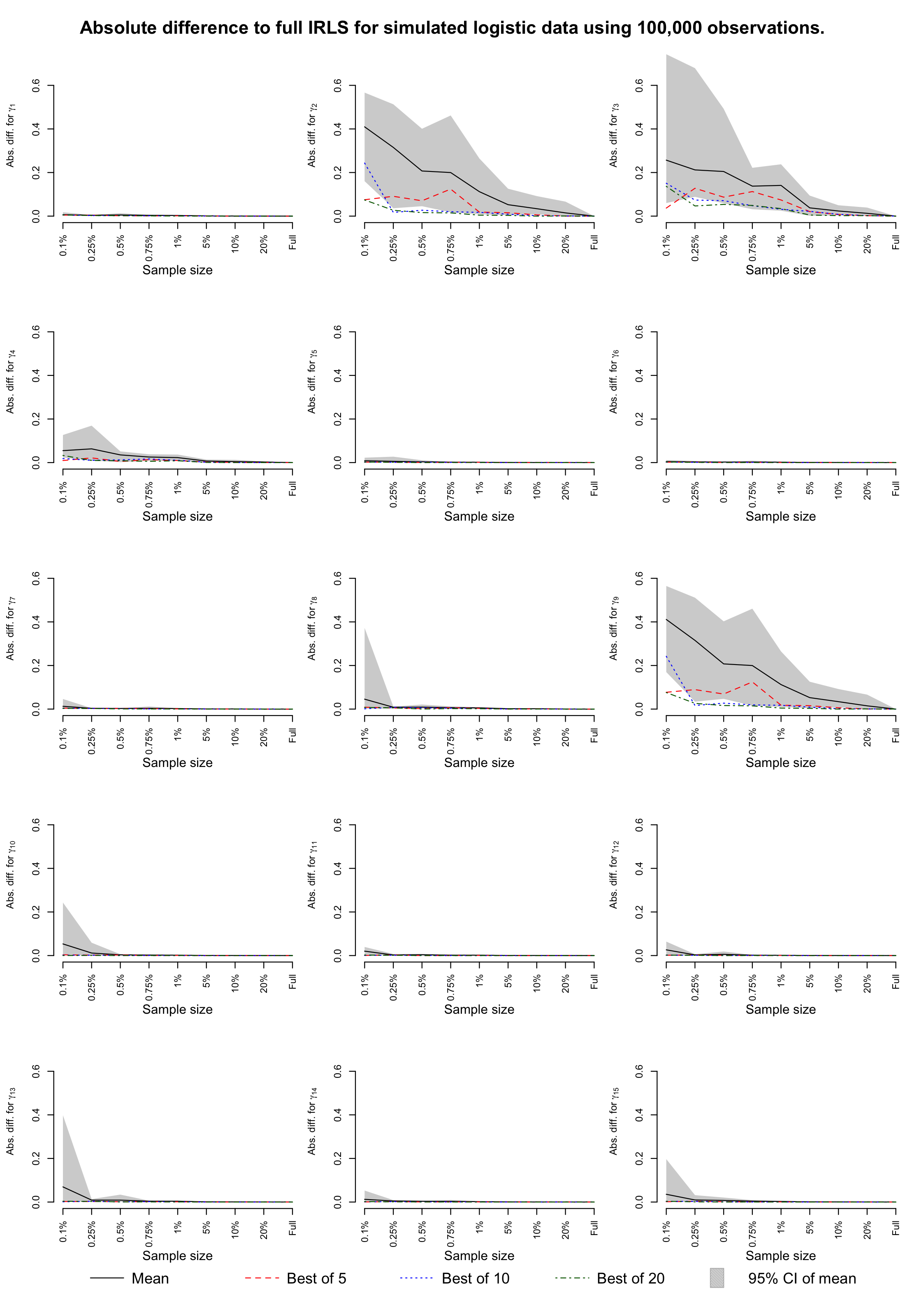}
	\caption{The absolute difference for marginal posterior probabilities for full enumeration of 32,768 logistic models using $100,000$ observations from the set of data in Example~\hyperref[example1]{1}. 20 runs were performed and the black line shows the mean absolute difference, with grey 90\% confidence intervals. The dashed and dotted lines show the absolute difference obtained by using the best likelihood estimate for each model that occurred during 5, 10 and 20 random runs.}
	\label{fig:full_100Kl}
\end{figure}

\FloatBarrier
\newpage

\section{Additional results for Example~\hyperref[example2]{2}}\label{A3}

\vspace{4.2cm}
\begin{figure}[h]
	\centering
	\includegraphics[scale=0.2]{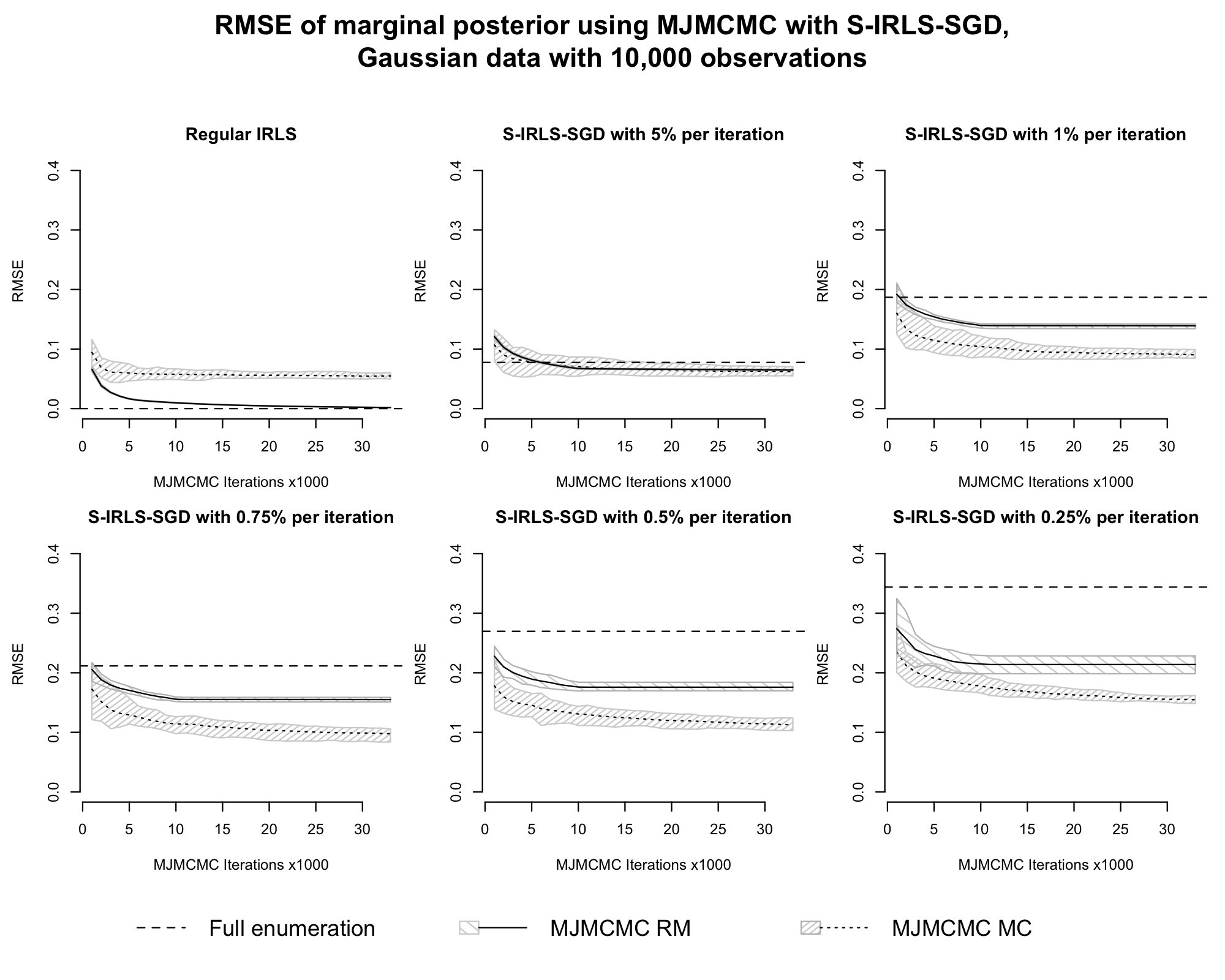}
	\caption{Root mean squared error compared to the true distribution for MJMCMC using both regular IRLS and S-IRLS-SGD with varying subsample sizes. MJMCMC was run 20 times and for 33,000 iterations for every variant of the algorithm. The black solid and dotted lines show the mean RMSE for the 20 runs, with the shaded areas showing 90\% confidence intervals, for RM and MC estimates respectively. The dashed line shows the RMSE for the full enumeration using the same algorithm settings.}
	\label{fig:mjmcmc_10Kg}
\end{figure}
\begin{figure}
	\centering
	\includegraphics[scale=0.2]{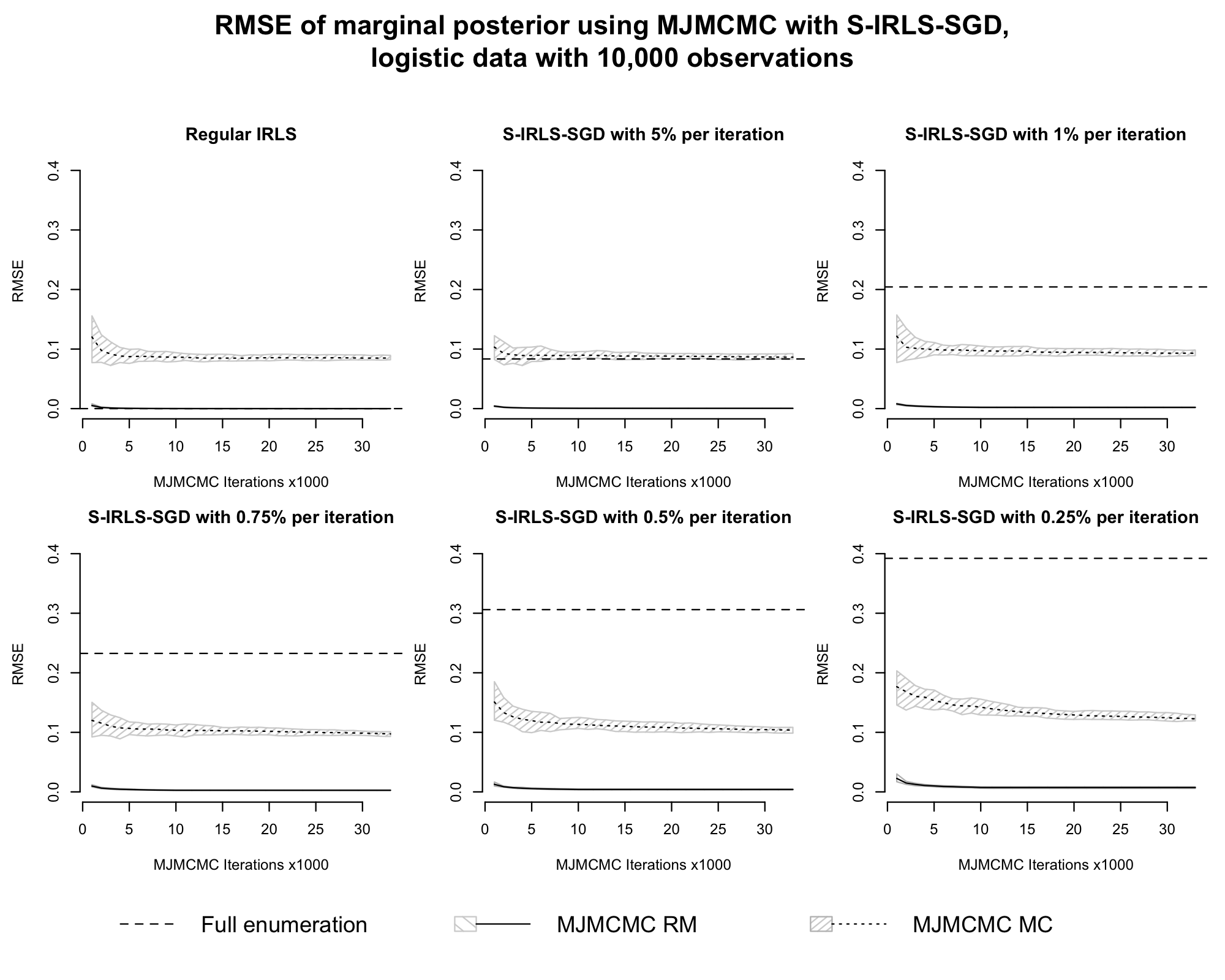}
	\caption{Root mean squared error compared to the true distribution for MJMCMC using both regular IRLS and S-IRLS-SGD with varying subsample sizes. MJMCMC was run 20 times and for 33,000 iterations for every variant of the algorithm. The black solid and dotted lines show the mean RMSE for the 20 runs, with the shaded areas showing 90\% confidence intervals, for RM and MC estimates respectively. The dashed line shows the RMSE for the full enumeration using the same algorithm settings.}
	\label{fig:mjmcmc_10Kl}
\end{figure}
\FloatBarrier






\end{document}